\documentclass[12pt]{iopart}
\usepackage{iopams}  
\usepackage[utf8]{inputenc}
\usepackage{cite} 
\usepackage{amsthm}
\usepackage{graphicx}
\usepackage{bbm}
\usepackage{hyperref}

\newcommand{\T}{\mathrm{T}}

\newcommand{\myThreeJ}[2]{
\Bigg(
\begin{array}{@{} c @{\hspace{2mm}} c @{\hspace{2mm}} c@{}}
  #1\\[-0.4mm]
  #2
\end{array}
\Bigg)
}

\newcommand{\delover}[2]{ \stackrel{#1}{\raisebox{-.2ex}{\normalsize$\square$}} \hspace{-0.8ex} (#2)  }
\newcommand{\del}[1]{ \square{(#1)} }
\newcommand{\xsub}[3]{{#1}^{\raisebox{#2}{\scriptsize$#3$}}} 
\newcommand{\overarrowethpower}[1]{\stackrel{\raisebox{-1.ex}{$#1$}}{ \ethpower{\s} }} 
\newcommand{\overarrowethadjpower}[1]{\stackrel{\raisebox{-1.ex}{$#1$}}{ \ethadjpower{\s} }} 
\newcommand{\overarroweth}[1]{\stackrel{\raisebox{0.25ex}{$#1$}}{ \eth }}
\newcommand{\overarrowethadj}[1]{\raisebox{-1.2ex}{$\stackrel{\raisebox{-1.ex}{$#1$}}{ \raisebox{1ex}{$\ethadj$} }$}} 
\newcommand{\ethpower}[1]{ \xsub{\eth}{0.5ex}{#1} } 
\newcommand{\ethadjpower}[1]{ \xsub{\ethadj}{0.ex}{#1} } 
\newcommand{\ethadj}{ \overline{\eth} }

\newcommand{\Proj}{\mathcal{P}}

\newcommand{\spinvacuumpure}{| J\!J \rangle }
\newcommand{\s}{ \eta }
\newcommand{\stars}{ \star^{(s)} }
\newcommand{\kappafactor}{ {}^{\s}\kappa_{jm,j'm'}^{\ell}}

\newcommand{\Y}{\mathrm{Y}}

\newcommand{\abs}[1]{\ensuremath{\vert #1 \vert}}
\newcommand{\pht}{| \alpha_0^+ \rangle}
\newcommand{\spinraised}{| \Omega_0^+ \rangle}
\newcommand{\vacuum}{| JJ\rangle}
\newcommand{\lp}{\mathcal{K}(\Omega_0)}
\newcommand{\lpc}{\overline{\mathcal{K}}(\Omega_0)}
\newcommand{\lpn}{\mathcal{K}}
\newcommand{\lpcn}{\overline{\mathcal{K}}}
\newcommand{\FF}{F_{K(\Omega_0)}}
\newcommand{\unity}{\mathbbmss{1}}
\newtheorem{definition}{Definition}

\newtheorem{theorem}{Theorem}

\newtheorem{proposition}[theorem]{Proposition}

\newtheorem{result}{Result}

\usepackage{xcolor}

\newcommand{\eprint}[1]{\href{https://arxiv.org/abs/#1}{arXiv:#1}}

\begin{document}

\title[Continuous phase spaces and the time evolution of spins]{Continuous phase spaces and the time evolution of spins:
star products and spin-weighted spherical harmonics}

\author{Bálint Koczor, Robert Zeier, and Steffen J. Glaser}

\address{Department Chemie, Technische Universität München,\\
Lichtenbergstrasse 4, 85747 Garching, Germany}

\begin{abstract}
We study continuous phase spaces 
of single spins and develop a complete description
of their time evolution. The time evolution
is completely specified by 
so-called star products.
We explicitly determine these star products for general spin numbers
using a simplified approach 
which applies spin-weighted spherical harmonics.
This approach naturally relates phase spaces of
increasing spin number 
to their quantum-optical limit
and allows for efficient approximations of the time evolution
for large spin numbers. We also
approximate phase-space representations
of certain quantum states that are
challenging to calculate for large spin numbers.
All of these applications are explored in concrete examples
and we outline extensions to coupled spin systems.
\end{abstract}

%
\vspace{2pc}

\section{Introduction}

Phase-space techniques provide a complete
description of quantum mechanics which is complementary to Hilbert-space \cite{cohen1991quantum}
and path-integral \cite{feynman2005} methods. These techniques are widely used
in order to describe, visualize, and analyze
quantum states  
\cite{Leonhardt97,carruthers1983,hillery1997,kim1991,lee1995,gadella1995,zachos2005,schroeck2013,SchleichBook,Curtright-review}.
Particular cases include Wigner \cite{wigner1932}, Husimi Q \cite{husimi1940},
and Glauber P  \cite{cahill1969} functions.
In this work, we are particularly interested in phase-space methods that are applicable to 
(finite-dimensional) spin systems
\cite{stratonovich,Agarwal81,DowlingAgarwalSchleich,
VGB89,brif97,brif98,heiss2000discrete,starprod,
klimov2002exactevolution,klimov2005classical,
klimov2008generalized,KdG10}
and how these methods
are related to infinite-dimensional phase spaces \cite{koczor2017}.
Building on earlier results in \cite{brif98,Agarwal81,DowlingAgarwalSchleich,heiss2000discrete,KdG10,tilma2016},
we have developed in \cite{koczor2017} a unified description
for the general class of $s$-parametrized phase spaces with $-1\leq s \leq 1$
which is applicable to single spins with integer or half-integer spin number $J$
and which naturally recovers the
infinite-dimensional case in the large-$J$ limit.
The $s$-parametrized phase-space function 
corresponding to a Hilbert-space operator $A$ is denoted by
$F_A (\Omega,s)$.

A new focus emerged recently with the objective to faithfully 
describe \emph{coupled} spin systems 
with the help of phase-space representations
\cite{PhilpKuchel,MJD,Harland,
GZG15,tilma2016,koczor2016,
rundle2017,Leiner17,Leiner18,
RTD17} while also emphasizing their spin-local properties.
In this context, we have completely characterized the time evolution of 
Wigner functions for
coupled spins $1/2$ in \cite{koczor2016} using  
explicit star products \cite{VGB89,starprod,koczor2016}.
Star products are an important concept in the phase-space description
of the time evolution and they
determine 
the phase-space 
function 
\begin{equation}
\label{firststarproddef}
F_{AB} (\Omega,s) = F_A (\Omega,s) \stars F_B (\Omega,s),
\end{equation}
of a product of two Hilbert-space operators
$AB$ in terms of the individual phase-space functions
$F_A (\Omega,s)$ and $F_B (\Omega,s)$ \cite{koczor2016}.
This results in the so-called Moyal equation
\begin{equation}
\label{moyaleq}
\frac{\partial  F_\rho (\Omega,s)}{\partial t}
=
-i \, F_{\mathcal{H}} (\Omega,s) \, \star^{(s)} F_\rho (\Omega,s)
+i \, F_\rho (\Omega,s) \, \star^{(s)} F_{\mathcal{H}} (\Omega,s)
\end{equation}
which describes time evolution of a quantum state $F_\rho (\Omega,s)$
under a Hamiltonian $F_{\mathcal{H}} (\Omega,s)$ directly in
phase-space (cf.\ \cite{koczor2016}).

\begin{figure}
	\centering
	\includegraphics[width=\textwidth]{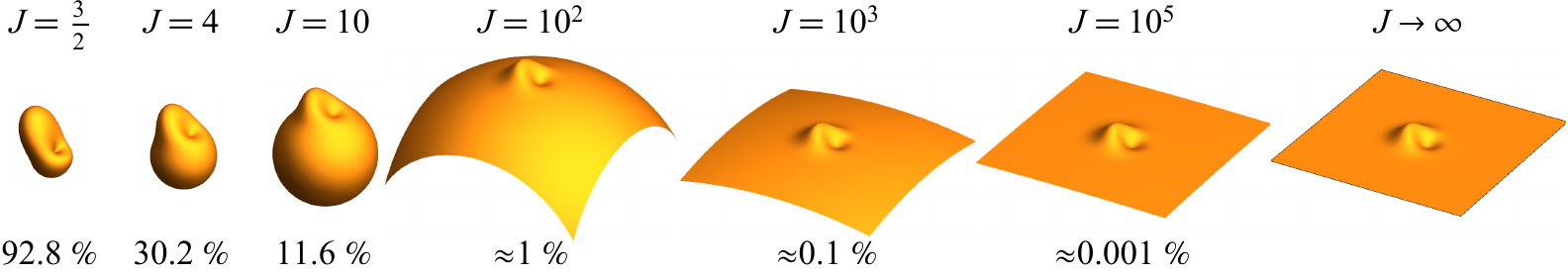}
	\caption{\label{convergencefig}
	         Example of how methods developed in this work 
	         can be applied (see \sref{examplesection}):
	         Efficiently computed approximations 
	         of spherical
		Wigner functions of the excited spin coherent state $\spinraised$
		(see \eref{exampledefinition}) are shown
		together with their $L^2$ errors. They concentrate 
		at the north pole  for increasing spin number $J$
		and approach the photon-added coherent state $\pht$
		(see \eref{infidimexample}),
		as the spherical phase space converges 
		to the planar one.
		The spherical Wigner functions are plotted on 
		a sphere of radius $R:=\sqrt{J/(2\pi)}$.
	}
\end{figure}

In this work, we extend our earlier results in \cite{koczor2016}
on Wigner functions of coupled spins $1/2$ 
and present the explicit form of the star product for 
the general class of $s$-parametrized phase spaces
which is applicable to
single and coupled spins of arbitrary spin number $J$.
We also rely on phase-space techniques for 
single spins $J$ that have been developed in \cite{koczor2017}.
We introduce spin-weighted spherical harmonics \cite{newman1966}
as an important new technical tool to the theory of phase spaces, 
even though they have not been considered in this context before.
This allows us to significantly simplify the theory 
of phase spaces and their star products.
In particular, we can now efficiently approximate
the time evolution of phase-space representations
for single spins that have a large spin number $J$.
Many quantum states have quite complicated 
phase-space representations which are challenging to
calculate for large values of $J$. Approximation methods
for the time evolution also lead to efficient 
computational techniques for approximating
phase-space representations of spins with large $J$
and \fref{convergencefig} illustrates this limit for Wigner functions 
of excited spin coherent states.
Relying on results from \cite{koczor2016}, we outline in the main text
how 
our results can be also extended to coupled spin systems.

Let us compare our work with earlier results.
The star product of Husimi Q functions of single spins has been derived
in \cite{kasperkovitz1990} using angular-momentum operators.
This approach has been independently rediscovered in \cite{starprod,klimov2002exactevolution}
and was used to calculate the star product and time evolution
of Glauber P functions, and this has also been translated 
to the general case of $s$-parametrized
phase-space representations.
A semiclassical equation of motion was derived in \cite{starprod}
by neglecting quantum terms of the star product, which 
has then been applied to the semiclassical simulation of
quantum dynamics  \cite{klimov2005classical} and
the classical limit of spin Bopp
operators \cite{zueco2007}.
In contrast to this approach
relying on angular momentum operators, our techniques
based on spin-weighted spherical harmonics and spin-weight raising and lowering operators
facilitate a simplified and more systematic approach and also lead to additional formulas
for the star product.
The derivation of the
exact star product is now completely transparent
and all of its quantum contributions are accounted for. 
One particular strength of our approach is that
the large-spin limit is naturally 
incorporated as spin-weight raising and lowering operators converge for large $J$ to
derivatives in the tangent plane (which are widely studied in infinite dimensions \cite{agarwal70II}).

This work has the following structure: 
We start in \sref{infdimrevieew} by recapitulating elementary properties of infinite-dimensional phase spaces
and their star products.
In \sref{findimreview},
we recall the structure of phase-space representations for
single spins $J$ following the approach of \cite{koczor2017}.
We introduce spin-weighted spherical harmonics
and summarize their main features
in \sref{spinweightedsphsection}.
Important approximation formulas for 
spin-weight raising and lowering operators 
are derived in \sref{ethapproximations}.
The sections~\ref{pqfunctionstarprod}-\ref{starproductssections}
constitute the main part of our work and various formulas for
exact and approximate star products are obtained.
Our methods are illustrated
with concrete examples
in sections~\ref{illustrativeexample}-\ref{examplesection}.
Before we conclude,
extension to coupled spin systems are outlined
in \sref{generalization}. Certain details and proofs are deferred to appendices.

\section{Phase spaces and star products in infinite dimensions \label{infdimrevieew}}
An important class of infinite-dimensional phase-space representations 
of a density operator $\rho$
contains $s$-parametrized 
phase-space distribution functions (where $-1 \leq s \leq 1$)
which can be defined via
\cite{koczor2017,cahill1969,moya1993,Leonhardt97,thewignertransform} 
\begin{equation}
\label{inifnitedimdefinition}
F_\rho (\Omega,s) = \mathrm{Tr}\,[ \, \rho \, \mathcal{D}(\Omega)  \Pi_s  \mathcal{D}^\dagger(\Omega) ].
\end{equation}
The distribution function $F_\rho (\Omega,s)$
is determined by the expectation value of the parity
operator $\Pi_s$ which is transformed by the displacement operator $\mathcal{D}(\Omega)$.
The parity operator inverts phase-space coordinates via
$\Pi_0 |\Omega \rangle = |{-} \Omega \rangle$ \cite{thewignertransform}
and the displacement operator $\mathcal{D}(\Omega)$ is defined by the property that it
translates 
the vacuum state $|0\rangle$ to coherent states $\mathcal{D} (\Omega) |0\rangle = |\Omega \rangle$.
Here, $\Omega$ parametrizes a phase space with either the variables $p$ and $q$ or
the complex eigenvalues $\alpha$ of the annihilation operator \cite{Leonhardt97}.
And the parameter $s$ interpolates between the Glauber P function for $s=1$ and the Husimi Q function for $s=-1$.
The particular case of $s=0$ corresponds to the Wigner function.
All $s$-parametrized phase-space distribution functions are
related to each other via Gaussian smoothing \cite{koczor2017,cahill1969},
and the convolution of 
the vacuum-state representation $F_{| 0 \rangle}(\Omega,s')$ 
with the distribution function $F_\rho(\Omega,s)$
results in a distribution function
\begin{equation}
\label{infinitedimensionalswitch}
F_\rho(\Omega,s{+}s'{-}1) = F_{| 0 \rangle}(\Omega,s')  \ast F_\rho(\Omega,s)
=
\exp[\case{1-s'}{2} \partial_{\alpha^*}\partial_{\alpha} ] \,  F_\rho(\Omega,s)
\end{equation}
of type $s{+}s'{-}1$.
The r.h.s.\ of \eref{infinitedimensionalswitch} establishes the corresponding differential form, see, e.g.,
Eq.~(5.29) in \cite{agarwal70}.

We adapt notations from \cite{brif98} for the infinite-dimensional tensor operators $\T_\nu := \mathcal{D}(\nu)$
which define the displacement operators using a continuous, complex index~$\nu$.
The phase-space representations $F_{\T_\nu} (\alpha,s) = \gamma_\nu^{-s} \, \Y_\nu(\alpha)$
are  up to the weight factor $\gamma_\nu:=\exp(- |\nu|^2/2)$ proportional
to the harmonic functions $\Y_\nu(\alpha) := \exp( \nu \alpha^* {-} \alpha \nu^*)$ \cite{brif98},
where the power $-s$ of $\gamma_\nu$ is determined by the type $s$ of the representation.
Up to a complex prefactor, multiplying two displacement operators results in a single displacement operator  \cite{cahill1969}:
\begin{equation}
\label{infdimdecomp}
\T_\mu  \T_\nu = \exp [ (\mu \nu^* {-} \nu \mu^*)/2] \; \T_{\mu + \nu}.
\end{equation}
Applying the product in \eref{infdimdecomp} and 
the star product from \eref{firststarproddef}, one obtains the 
formula
\begin{equation}
\label{infdimstarprod}
[\gamma_\mu^{-s} \, \Y_\mu(\alpha) ] \stars [\gamma_\nu^{-s} \, \Y_\nu(\alpha)] 
= \exp [ (\mu \nu^* {-} \nu \mu^*)/2] \,  \gamma_{\mu {+} \nu}^{-s} \,  \Y_{\mu {+} \nu}(\alpha).
\end{equation}
The star product satisfies
 \eref{infdimstarprod} and it can be explicitly defined as a power series 
\begin{equation}
\label{infinitedimstarprod}
\stars := \exp [ \case{(1{-}s)}{2} \overleftarrow{\partial}_{\alpha} \overrightarrow{\partial}_{\alpha^*} 
-
\case{(1{+}s)}{2} \overleftarrow{\partial}_{\alpha^*} \overrightarrow{\partial}_{\alpha}]
\end{equation}
of partial derivatives as in Eq.~(3.5) of \cite{agarwal70II}.
Setting $\alpha = (\lambda q {+} i \lambda^{-1} p)/\sqrt{2}$ for arbitrary
real $\lambda$ \cite{cahill1969}, the derivatives observe (see also Eq.~(3.4') in \cite{agarwal70II})
\begin{equation*}
	\fl \qquad 
\case{(1{-}s)}{2}   \overleftarrow{\partial}_{\alpha}  \overrightarrow{\partial}_{\alpha^*} 
-
\case{(1{+}s)}{2} \overleftarrow{\partial}_{\alpha^*} \overrightarrow{\partial}_{\alpha}
=
i[
\overleftarrow{\partial}_q \overrightarrow{\partial}_p - \overleftarrow{\partial}_p \overrightarrow{\partial}_q
-s \lambda^{2} \overleftarrow{\partial}_p \overrightarrow{\partial}_p
-s \lambda^{-2} \overleftarrow{\partial}_q \overrightarrow{\partial}_q
]/2.
\end{equation*}
The arrows represent whether derivatives are to be taken to the left or right.
For $s=0$, we obtain $\star^{(0)}=\exp( i \{\cdot,\cdot\} /2)$
as stated by Groenewold \cite{Gro46} and
$\{\cdot,\cdot\}=\overleftarrow{\partial}_q \overrightarrow{\partial}_p - \overleftarrow{\partial}_p \overrightarrow{\partial}_q$
denotes the Poisson bracket.

\section{Finite-dimensional phase spaces \label{findimreview}}

We briefly review $s$-parametrized phase-space representations
for a single spin with spin number $J$ following the approach of \cite{koczor2017},
which recovers
the previously discussed infinite-dimensional case in the large-spin limit. 
The continuous phase space  for 
the finite-dimensional spin $J$ is fully parametrized using
the two Euler angles $\Omega := (\theta,\phi)$ of the rotation operator
$\mathcal{R}(\Omega)=\mathcal{R}(\theta,\phi):= e^{i\phi \mathcal{J}_z} e^{i\theta \mathcal{J}_y} $.
Here, $\mathcal{J}_z$ and $\mathcal{J}_y$ are components of the 
angular momentum operator that are  defined by their commutation
relations, i.e., $[\mathcal{J}_j,\mathcal{J}_k]=i \sum_\ell \epsilon_{jk\ell} \mathcal{J}_\ell$
where $j,k,\ell \in \{x,y,z\}$ and $\epsilon_{jk\ell}$ is the Levi-Civita symbol
\cite{messiah1962}.
This leads to a spherical phase space with
radius $R:=\sqrt{J/(2\pi)}$. The displacement operator
$\mathcal{D}(\Omega)$ from the infinite-dimensional case
is replaced by the rotation operator
$\mathcal{R}(\Omega)$
which maps the spin-up state $\spinvacuumpure$ to spin coherent states
$| \Omega \rangle = \mathcal{R}(\Omega) \spinvacuumpure$
\cite{perelomov2012,arecchi1972atomic,DowlingAgarwalSchleich}.
The $s$-parametrized phase-space representation
\begin{equation}
\label{PSrepDefinition}
F_\rho (\Omega,s) :=  \Tr \,[ \, \rho \, \mathcal{R}(\Omega)  M_s  \mathcal{R}^{\dagger}(\Omega) ]
\end{equation}
of a density operator $\rho$ of a single spin $J$
is then obtained as the expectation value of the rotated (generalized) parity operator
(refer to \cite{koczor2017})
\begin{equation}
\label{DefOfspinParityoperators}
M_s := \case{1}{R} \, \sum_{j=0}^{2J} \sqrt{\case{2j{+}1}{4 \pi}} (\gamma_j)^{-s} \, \T_{j0},
\end{equation}
where $M_s$ is specified in terms of a
weighted sum of tensor operators $\T_{jm}$ of order zero (i.e., $m=0$).
The tensor components $\T_{jm}$
depend on the rank $j\in\{0,\ldots,2J\}$, the order $m\in\{-j,\ldots,j\}$, 
and the spin number $J$.
The explicit matrix elements
are specified 
in terms of
Clebsch-Gordan coefficients \cite{messiah1962,brif98,BL81,Fano53} as
\begin{equation}
[\T_{jm}]_{m_1 m_2} :=  \sqrt{\case{2j{+}1}{2J{+}1}} \, C^{J m_1}_{J m_2, j m}
=(-1)^{J-m_2}\, C^{jm}_{Jm_1J,-m_2},
\end{equation} 
where $m_1,m_2 \in \{J,\ldots,-J\}$.
The weight factor has the explicit form
\begin{equation}
\label{gammafactor}
\gamma_j:=R\, \sqrt{4\pi} (2J)! \, [ (2J{+}j{+}1)! \, (2J{-}j)! \,  ]^{{-}1/2},
\end{equation}
and the power $-s$ of $\gamma_j$ determines the type $s$ of the phase-space representation \cite{koczor2017}.

Tensor operators $\T_{jm}$ form an orthonormal basis of  $(2J{+}1) \times  (2J{+}1)$
matrices with respect to the Hilbert-Schmidt scalar product
$\mathrm{Tr}\,[ \, \T_{jm}   \T_{j'm'}^\dagger ] = \delta_{j j'}\delta_{m m'}$
where $ 0 \leq j \leq 2J$ and $m,m'\in\{-j,\ldots,j\}$.
Similarly as in \eref{infdimdecomp}, the product of two tensor operators can be decomposed into
a sum (applying the notation of \cite{koczor2016})
\begin{equation}
\label{topdecomposition}
\T_{jm}  \T_{j'm'} = \sum_{\ell=0}^{2J} K_{jm,j'm'}^{\ell} \T_{\ell, m + m'}
\end{equation}
of tensor operators using
the decomposition coefficients $K_{jm,j'm'}^{\ell}$ as detailed in \ref{productexpansionappendix}.
The phase-space representations 
$F_{\T_{jm}} (\Omega,s) = \gamma_j^{-s} \, \Y_{jm}(\Omega) /R$
of tensor operators are proportional 
to spherical harmonics of rank $j$ and order $m$ and they are orthonormal
with respect to a spherical integration
\begin{equation} \label{integral}
\int_{S^2} \Y_{jm}(\Omega) \Y^*_{j'm'}(\Omega) / R^2 \, \mathrm{d} \Omega  = \delta_{j j'}\delta_{m m'}
\end{equation}
according to $\mathrm{d} \Omega = R^2 \sin{\theta} \mathrm{d} \theta \, \mathrm{d} \phi$.
Similarly as in \eref{infdimstarprod}, the defining property of the star product
for phase-space representations of spins can be transferred to spherical harmonics since
these distribution functions are always given as a finite linear combination of spherical harmonics:

\begin{definition}
	\label{def1}
As in \eref{firststarproddef},
the star product $\stars$ of two phase-space representations of type $s$ satisfies
for a single spin $J$
the condition
\begin{equation}
\label{starproductcondition}
[\gamma_j^{-s} \, \Y_{jm}(\Omega)] \stars [ \gamma_{j'}^{-s} \, \Y_{j'm'}(\Omega)]
=
R \sum_{\ell=0}^{2J} K_{jm,j'm'}^{\ell} \, \gamma_\ell^{-s} \, \Y_{\ell, m + m'}(\Omega)
\end{equation}
for all suitable indices with $j,j'\leq 2J$. The coefficients $K_{jm,j'm'}^{\ell}$ are  determined by
\eref{topdecomposition}.
\end{definition}
One objective of this work is to apply this decomposition in order to define 
a star product $\stars$ in terms of \emph{spin-weighted}
spherical harmonics and their spin-weight raising and lowering 
differential operators $\eth$ and $\ethadj$.

\section{Spin-weighted spherical harmonics \label{spinweightedsphsection}}
Spin-weighted spherical harmonics $\Y^{\s}_{j m}$ with spin weight $\s$
have been introduced by Newman and Penrose \cite{newman1966} using 
spin-weight raising  and lowering operators $\eth$ and $\ethadj$
in order to describe the asymptotic behavior of the gravitational field.
Since then, spin-weighted spherical harmonics have been widely used 
to analyze, in particular,
gravitational waves \cite{thorne1980,seljak1997} or the cosmic microwave
background \cite{zaldarriaga1997,okamoto2003}.
Moreover, efficient computational tools for spin-weighted spherical harmonics
are available  \cite{libsharp} and these can also be used in
the fast calculation of spherical convolutions \cite{beamdeconv,fastconv}.

\begin{figure}
	\centering
	\includegraphics{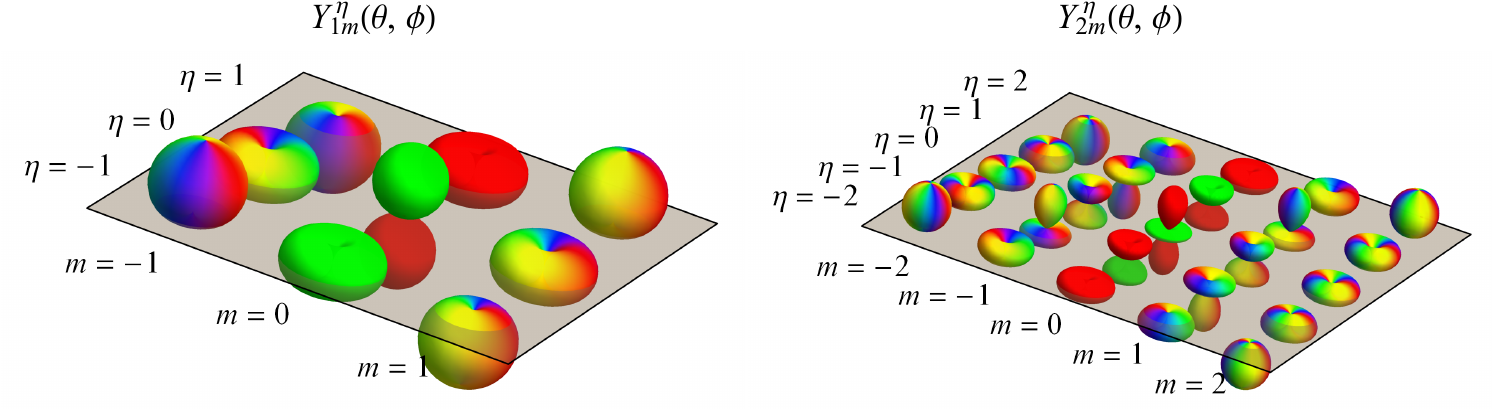}
	\caption{Spin-weighted spherical harmonics $\Y^{\s}_{j m}$ of rank
		one (left) and two (right). Colors represent the complex phase;
		red (dark gray) and green (light gray) depict
		positive and negative values, while blue and yellow
		represent $i$ and ${-}i$.
		\label{fig:spin_weight} }
\end{figure}

Spin-weighted spherical harmonics $\Y^{\s}_{j m}$
with $-j \leq \s \leq j$ are defined
as functions on the three-dimensional sphere as shown in \fref{fig:spin_weight}.
Similarly as for ordinary spherical harmonics $\Y_{j m}(\theta,\phi)$,
the spin-weight raising  and
lowering  operators $\eth$ and $\ethadj$ are used in their definition 
(see, e.g., \cite{newman1966} and Chapter~2.3 in \cite{Del2012})
\begin{equation}
\label{SPHdefininition}
\Y^{\s}_{j m} :=
\cases{
\hspace{10mm} 	\sqrt{{(j{-}\s)!}/{(j{+}\s)!}} \; \ethpower{\s} \, \Y_{jm} 
&for $\s \geq 0$
	\\
(-1)^\s \sqrt{{(j{-}|\s|)!}/{(j{+}|\s|)!}} \; \ethadjpower{|\s|} \, \Y_{jm} 
&for $\s<0$,
\\}
\end{equation}
and the particular case of $\s=0$ corresponds to ordinary spherical harmonics.
(Spin-weighted spherical harmonics are related to Wigner D-matrices via
$
D_{m \s}^{j}(\phi,\theta,\psi)=(-1)^{m} \sqrt{(4\pi) /  (2j{+}1)}\Y^\s_{j, -m}(\theta,\phi)   e^{-i \s \psi}$,
refer to Eq.~(2.52) in \cite{Del2012}.)
The operators $\eth$ and $\ethadj$ raise and lower
the spin weight $\s$ with $-j \leq \s \leq j$ in (see Eq.~(3.20) in \cite{newman1966})
\begin{equation} \label{ETHoperatorC}
\fl \qquad \qquad
\eth \Y^{\s}_{j m} = \sqrt{(j{-}\s)(j{+}\s{+}1)} \; \Y^{\s+1}_{j m}
\;
\textup{ and }
\;
\ethadj \Y^{\s}_{j m} = -\sqrt{(j{+}\s)(j{-}\s{+}1)} \; \Y^{\s-1}_{j m}.
\end{equation}
Their explicit form can be specified in terms of the differential operators
\begin{eqnarray} \label{ETHoperatorA}
& \eth \Y^{\s}_{j m} =  - (\sin{\theta})^\s 
( \partial_\theta + i/\sin{\theta} \; \partial_\phi )
[ \, (\sin{\theta})^{-\s} \; \Y^{\s}_{j m}   ] \; \textup{ and}
\\
\label{ETHoperatorB}
& \ethadj \Y^{\s}_{j m} =  - (\sin{\theta})^{-\s}
( \partial_\theta - i/\sin{\theta} \; \partial_\phi )
[ \, (\sin{\theta})^{\s} \; \Y^{\s}_{j m} ],
\end{eqnarray}
see Eq.~(3.8) in \cite{newman1966}.
Spin-weighted spherical harmonics are up to a constant factor invariant under the application of
$\eth\ethadj$  and $\ethadj\eth$ (see Eq.~(2.22) in \cite{Del2012}):
\begin{equation}
\label{eigenfunct}
\fl \qquad \qquad
\ethadj  \eth  \Y^{\s}_{j m} = [\s (\s{+}1) - j  (j{+}1)] \; \Y^{\s}_{j m}
\;
\textup{ and }
\;
\eth \ethadj \Y^{\s}_{j m} =  [\s (\s{-}1) - j  (j{+}1)] \; \Y^{\s}_{j m}.
\end{equation}
Therefore, $\eth \ethadj $ acts up to a minus sign as the total angular momentum operator
when applied to spherical harmonics, i.e., $\eth \ethadj \Y_{j m}  = -j(j{+}1)\Y_{j m}$,
refer to Eq.~(2.25) in \cite{Del2012}.
The commutator $[\ethadj,\eth] = 2\s$ immediately follows from \eref{eigenfunct}.
Products $\Y^{\s}_{j m} \Y^{-\s}_{j' m'}
\propto (\ethpower{\s} \, \Y_{j m}) (\ethadjpower{\s} \, \Y_{j m})$
of spin-weighted spherical harmonics
decompose into the sums  \cite{Del2012}
\begin{eqnarray}
\label{spinweighteddecompositionA}
(\ethpower{\s} \,  \Y_{j m}) (\ethadjpower{\s} \, \Y_{j' m'})
&=  
\sum_{\ell=0}^{j+j'} \kappafactor \, \Y_{\ell, m + m'}(\Omega) \; \textup{ and} \\
\label{spinweighteddecompositionB}
(\ethadjpower{\s} \,  \Y_{j m}) (\ethpower{\s} \, \Y_{j' m'})
&=  
\sum_{\ell=0}^{j+j'} {}^{-\s}\kappa_{jm,j'm'}^{\ell} \, \Y_{\ell, m + m'}(\Omega)
\end{eqnarray}
of spherical harmonics.
The decomposition coefficients $\kappafactor$ are explicitly specified in \eref{kappaexplicit}
and they are similar to the ones used in Definition~\ref{def1}.
In sections \ref{pqfunctionstarprod}-\ref{starproductssections}, we utilize the products of spin-weighted spherical harmonics
from \eref{spinweighteddecompositionA} and \eref{spinweighteddecompositionB} 
to explicitly  determine the star product such that it satisfies its defining property from \eref{firststarproddef}.

\section{Approximating spin-weight raising  and lowering operators \label{ethapproximations}}

\begin{figure}
	\centering
	\includegraphics[width=\textwidth]{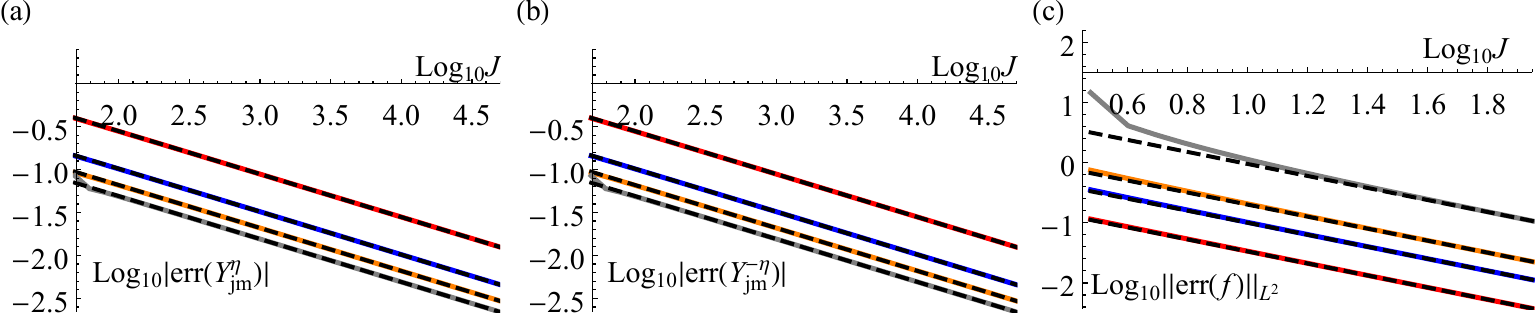}
	\caption{\label{approxfig} The absolute value of the difference between exact spin-weighted spherical
		harmonics (a) $\Y^{\s}_{j m} (\alpha)$ and (b) $\Y^{-\s}_{j m} (\alpha)$
		and their approximations from \eref{SPHapproxA} and \eref{SPHapproxB} for fixed $\s=4$, $m=4$, and
		$\alpha = 1.2 e^{i 2.1}$ and variable $j\in\{4,8,12,16\}$ (red, blue, orange, gray).
		(c) $L^2$-norm of the difference between the exact derivative of the spin-up state
		Wigner function	$f:=[\ethadj / \sqrt{2J} ]^n \, W_{\spinvacuumpure}(\Omega)$ and its approximation
		using Proposition~\ref{proposition1} for $n\in\{2,3,4,8\}$ (red, blue, orange, gray).
		Dashed lines show the expected scaling $\mathcal{O}( J^{-1/2})$ for (a) and (b) as well as
		$\mathcal{O}( J^{-1})$ for (c)
		with suitable prefactors.
	}
\end{figure}

As mentioned in \sref{findimreview}, the spherical phase space converges 
for an increasing spin number $J$ to the (infinite-dimensional) planar phase space
\cite{koczor2017}.
The arc length $\theta  R$ becomes a measure of distance from the north pole, which is
equivalent to its infinite-dimensional counterpart $\abs{\alpha}$. 
The two phase spaces can be related using the formula $\alpha = \sqrt{J/2} \, \theta \, e^{{-}i\phi}$.
In this parametrization,
spin-weighted spherical harmonics can, up to an additive error 
that scales inversely with $R$,
be expanded as derivatives
\begin{eqnarray}  \label{SPHapproxA}
\Y^{\s}_{j m} (\alpha)  
&=
(-1)^{\s} e^{ - i \s \phi} ({\partial}_{\alpha^*})^{\s} \, \, \Y_{j m}(\alpha)
+  \mathcal{O}( |\alpha| / \sqrt{J}) \;\textup{ and}\\   \label{SPHapproxB}
\Y^{-\s}_{j m}  (\alpha)
&=
e^{ i \s \phi} ({\partial}_{\alpha})^{\s} \, \, \Y_{j m}(\alpha)
+  \mathcal{O}( |\alpha| / \sqrt{J})
\end{eqnarray}
of ordinary spherical harmonics
with respect to the coordinates $\alpha$ and $\alpha^*$
while assuming a
fixed arc length $|\alpha|$.
This is essentially an approximation of \eref{SPHdefininition} for small angles $\theta$.
\Fref{approxfig}(a)-(b) plots the absolute value of the difference between the spin-weighted spherical
harmonics $\Y^{\s}_{j m} (\alpha)$ and their approximations which rely on the derivatives from \eref{SPHapproxA}-\eref{SPHapproxB}.
The approximation error vanishes in the limit of large $J$ assuming that the
coordinates $\alpha$ are located at the north pole or that the values $|\alpha| / \sqrt{J}$ are small
(e.g., $|\alpha|$ is bounded).
Extending this to the spin-weight raising and lowering operators 
from \eref{ETHoperatorA} and \eref{ETHoperatorB},
these operators $\eth$ and $\ethadj$ can be shown to transform  
for large $J$ into the derivatives ${\partial}_{\alpha^*}$ and ${\partial}_{\alpha}$ 
over the complex plane.
\begin{proposition}
	\label{proposition1}
	Assume that the phase-space function $f= F_\rho(\Omega,s)$
	of a spin $J$ is parametrized using the arc length $\alpha = \sqrt{J/2} \, \theta \, e^{{-}i\phi}$
	and that its spherical-harmonics expansion coefficients might depend on $J$.
	The action of spin-weight raising and lowering  operators 
	$\eth$  and $\ethadj$ at fixed $\alpha$ are given by 
	\begin{eqnarray}
	[(\eth / \sqrt{2J} )^{\s} \, f](\alpha)   & =  (-1)^{\s} e^{- i \s \phi} ({\partial}_{\alpha^*})^{\s} \, f(\alpha)
	+
	 \mathcal{O}(|\alpha| J^{-1}),
	\\ \hspace{0mm}
	[( \ethadj / \sqrt{2J} )^{\s} \, f](\alpha)  & = 
	(-1)^{\s} e^{ i \s \phi} ({\partial}_{\alpha})^{\s} f(\alpha)
	+
	\mathcal{O}(|\alpha| J^{-1}), \\ \hspace{0mm}
        [ (\,  \eth \ethadj /(2J) 	\,) f](\alpha) & = {\partial}_{\alpha^*}{\partial}_{\alpha} \,f(\alpha)  + \mathcal{O}(|\alpha| J^{-1}), 
        \; \textup{ and}
        \\ \hspace{0mm}
	[ (\, \ethadj \eth /(2J)  \,) f](\alpha) & = {\partial}_{\alpha} {\partial}_{\alpha^*} \,f(\alpha) + \mathcal{O}(|\alpha| J^{-1}),
	\end{eqnarray}
	and this action is up to an error term $\mathcal{O}(J^{-1})$  equivalent
	to applying the complex derivatives ${\partial}_{\alpha^*}$ and ${\partial}_{\alpha}$
	for any powers of $\s$.
	The error term $\mathcal{O}(J^{-1})$ vanishes in the limit of large  $J$ if the
	differentials $[\eth / \sqrt{2J} ]^\s \, f$ remain non-singular in the limit.
	This implies convergence in the $L^2$ norm if $f$
	and its differentials are also square integrable in the limit.
	Refer to \ref{diffopconvergence} for the proof.
\end{proposition}

Note that phase-space representations $F_\rho(\Omega,s)$ and all their derivatives are non-singular
and square integrable if $\rho$ is finite-dimensional, i.e., if $F_\rho(\Omega,s)$ is a finite
linear combination of spherical harmonics. In general,
singularities can however appear for $s>0$
in the limit of large $J$, as the corresponding parity operators $\Pi_s$ from
\eref{inifnitedimdefinition} are unbounded \cite{cahill1969,koczor2017} for $s>0$.
This can be illustrated using the example of the spin-up state $F_{\spinvacuumpure}(\Omega,s)$:
it is determined by a sum of $2J{+}1$ spherical harmonics and 
its expansion coefficients are proportional to $\gamma_j^{1-s}$ and
depend implicitly on $J$ \cite{koczor2017}.
These rapidly decreasing expansion coefficients can be approximated 
for increasing $j$ by
$e^{-j^2(1-s)/(4J)}$  if $s<1$ and
$F_{\spinvacuumpure}(\Omega,s)$ is bounded and square integrable in the large-$J$ limit.
But for $s=1$ this expansion defines in the limit 
a delta distribution which is clearly singular and not square integrable.
For example, the differentials $[\ethadj / \sqrt{2J} ]^{\s} F_{\spinvacuumpure}(\Omega,s)$ are sums of spin-weighted spherical harmonics
with expansion coefficients which are proportional to $[(2J){(j{-}\s)!}/{(j{+}\s)!}]^{-\s/2} \,  \gamma_j^{1-s}$
and which can be approximated by $[j^2/(2J)]^{\s/2} e^{-j^2(1-s)/(4J)}$. The coefficients vanish for increasing $j$ and define bounded,
square-integrable functions in the limit of large $J$ for $s < 1$. \Fref{approxfig}(c) shows the $L^2$ norm of the difference
between the Wigner function's differential $[\ethadj / \sqrt{2J} ]^{\s} \, W_{\spinvacuumpure}(\Omega)$ and its approximation 
via Proposition~\ref{proposition1}. This difference vanishes for large  $J$.
Refer to \ref{diffopconvergence} for further details.

One example of an unbounded operator is the Wigner function $W_{\mathcal{I}^+/\sqrt{2J}}(\Omega)$
of the raising operator $\mathcal{I}^+/\sqrt{2J}$ which reproduces
the annihilation operator $a$ in the large-spin limit \cite{arecchi1972atomic}.
One obtains
\begin{eqnarray}
&W_{\mathcal{I}^+/\sqrt{2J}}(\Omega) \propto  \sqrt{J/2} \,  \Y_{1,1}(\theta,\phi)
\propto \sqrt{J/2} \, \sin\theta e^{i \phi} \;\textup{ and}\\
&\ethadj/\sqrt{2J} \, W_{\mathcal{I}^+/\sqrt{2J}}(\Omega) \propto  \case{1}{2}  \Y^{-1}_{1,1}(\theta,\phi)
\propto \case{1}{2} (1+\cos\theta) e^{i \phi}.
\end{eqnarray}
The corresponding $L^2$ norms (as defined with respect to \eref{integral}) diverge with increasing $J$ and
the Wigner functions are no longer square integrable. However, for any bounded
$\alpha = \sqrt{J/2} \, \theta \, e^{{-}i\phi}$, the functions have the proper limits
\begin{equation*}
\fl \qquad \quad
\lim_{J \rightarrow \infty } [ \,  \sqrt{J/2} \, \sin(|\alpha|/\sqrt{J/2}) e^{i \phi} \, ] = \alpha  
\; \textup{ and } \;
\lim_{J \rightarrow \infty } [ \, 
\case{1}{2} (1+\cos(|\alpha|/\sqrt{J/2})
] = 1,
\end{equation*}
where $1$ is the derivative of $\alpha$. Refer to \ref{diffopconvergence} for details.

Proposition~\ref{proposition1} essentially approximates the differentials
$\eth^{\s} F_\rho(\Omega,s)$ of spherical functions with the derivatives
${\partial}_{\alpha^*} F_\rho(\Omega,s)$. This duality then becomes exact in
the large-spin limit if the differentials $\eth^{\s} F_\rho(\Omega,s)$ remain non-singular.
In particular, we use the spin-weight raising and lowering operators to construct the star product
by applying \eref{starproductcondition}, which then naturally recovers the infinite-dimensional
star product from \eref{infinitedimstarprod}:
\begin{proposition} \label{proposition2}
	Consider the arc-length parametrization $\alpha = \sqrt{J/2} \, \theta \, e^{{-}i\phi}$
	and two spin-$J$ phase-space functions $f= F_\rho(\Omega,s)$ and $g= F_{\rho'}(\Omega,s)$.
	Their spherical-harmonics expansion coefficients might depend on $J$.
	Following Proposition~\ref{proposition1}, one obtains the approximations
\begin{eqnarray} 
	f  [ \, 
	(\overarrowethpower{\leftarrow} ) 
	(\overarrowethadjpower{\rightarrow}) / (2J)^\s
	\, ] g
	& =
	f  [ \, (\overleftarrow{\partial}_{\alpha^*})^\s (\overrightarrow{\partial}_{\alpha})^\s \, ] g
	+ \mathcal{O}( J^{-1}), \\
	f [ \, 
	(\overarrowethadjpower{\leftarrow})
	(\overarrowethpower{\rightarrow} )  / (2J)^\s
	\,] g
	& =
	f [\, (\overleftarrow{\partial}_{\alpha})^\s
	(\overrightarrow{\partial}_{\alpha^*})^\s \, ] g 	+ \mathcal{O}(J^{-1}).
\end{eqnarray}
The error vanishes in the limit of large $J$ if the differentials 
$[\ethpower{\s} f] \, [\ethadjpower{\s} g]/(2J)^{\s}$ remain non-singular in the limit.
\end{proposition}

\section{Star products of spin Glauber P and Husimi Q functions \label{pqfunctionstarprod}}
\subsection{The exact star product}
We start by determining the exact star product of Q functions ($s{=}-1$)
and P functions ($s{=}1$) which are given uniquely in terms of
spin-weight raising and lowering operators:

\begin{result} \label{res1}
The (finite-dimensional) exact star product of two  Q functions $Q_A$
and $Q_B$ and two P functions $P_A$ and $P_B$ is determined for the spin number $J$ by 
\begin{equation} \label{result1exateq}
\fl \; \;
Q_A \star^{(-1)} Q_B 
= 
Q_A   \sum_{\s = 0}^{2J} 
\lambda^{(-1)}_\s
(\overarrowethadjpower{\leftarrow}) 
(\overarrowethpower{\rightarrow} )
 Q_B
\; \textup{ and }\;
P_A \star^{(1)} P_B
=   
P_A  \sum_{\s = 0}^{2J} 
\lambda^{(1)}_\s
(\overarrowethpower{\leftarrow} ) 
(\overarrowethadjpower{\rightarrow}) 
P_B, 
\end{equation}
where the coefficients (see \ref{proofofresult1})
\begin{equation} \label{exactstarprodceffs}
\lambda^{(-1)}_\s = \frac{(2J{-}\s)! }{\s! (2J)!}
\; \textup{ and } \; 
\lambda^{(1)}_\s = \frac{ R^2 4\pi \,  (-1)^\s  \,  (2J)! }{ \s!(2J{+}\s{+}1)!}
\end{equation}
depend on $J$. 
Terms in \eref{result1exateq}
related to spherical harmonics $\Y_{jm}$
with rank $j > 2J$ are not relevant 
and can be projected out as detailed in
\ref{Vandermonde} and \cite{koczor2016}.
\end{result}
The proof of Result~\ref{res1} is given in \ref{proofofresult1}.
The upper bounds in the sums in \eref{result1exateq} can be lowered to $\min{(j_A,j_B)}$
where $j_A$ and $j_B$ are the maximal ranks in the tensor-operator decompositions 
of $A$ and $B$. 
Similar results have been attained for Q functions in Eq.~(86) of \cite{kasperkovitz1990}
using angular momentum operators, refer also to Eq.~(45)
in \cite{starprod}. We improve and simplify these results 
for star products of phase-space representations of spins
with the help of  spin-weighted spherical harmonics.
In particular, this approach enables us to efficiently approximate 
phase-space representations for large spin numbers $J$
as discussed below.

\begin{figure}[b]
	\centering
	\includegraphics[width=\textwidth]{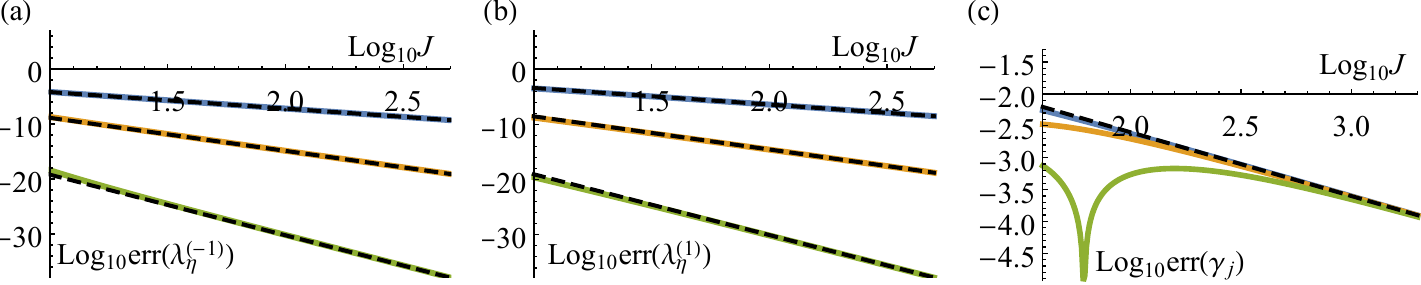}
	\caption{\label{coeffapproxfig} 
		(a)-(b) Difference between the exact factors $\lambda^{(-1)}_\s$ and $\lambda^{(1)}_\s$
		from Result~\ref{res1} \eref{exactstarprodceffs} and their approximations from
		\eref{lambdaapprox1} and \eref{lambdaapprox2}
		for $\s\in\{2,5,10\}$ (blue, orange, green).
		(c) Difference between the exact factor $\gamma_j$ and its approximation from Result~\ref{res4}
		for $j\in\{2,5,12\}$ (blue, orange, green).
		Dashed lines show the expected scaling $\mathcal{O}(J^{-\s-1})$ for (a) and (b) 
		as well as $\mathcal{O}(J^{-1})$ for (c)
		with suitable prefactors.
	}
\end{figure}

\subsection{Approximations of the star product \label{sec:approx_star}}

The coefficients in \eref{exactstarprodceffs} of Result~\ref{res1} can be expanded as (see \ref{gammaconvergence})
\begin{eqnarray} 
\label{lambdaapprox0}
\lambda^{(-1)}_\s  &= [\s! (2J)^{\s}]^{-1} \; \textup{ for $\s=0$ or $\s=1$,}\\
\label{lambdaapprox1}
\lambda^{(-1)}_\s  &= [\s! (2J)^{\s}]^{-1}  + \mathcal{O}(J^{-\s-1}) \; \textup{ for $\s \geq 2$, and}\\
\label{lambdaapprox2}
\lambda^{(1)}_\s  &=  (-1)^\s [\s! (2J)^{\s}]^{-1}  + \mathcal{O}(J^{-\s-1}) \; \textup{ for $\s \geq 0$}.
\end{eqnarray}
Also, the finite sums in \eref{result1exateq}
are unchanged if higher-order differentials (with respect to $\s$) are added since all the higher-order differentials
vanish, i.e., $\ethpower{\s} \Y_{j m}=0$ for $\s> j$. 
The scaling $\mathcal{O}(J^{-\s-1})$ of the approximations in \eref{lambdaapprox1}
and \eref{lambdaapprox2}
is  highlighted in \fref{coeffapproxfig}(a)-(b)
for different values of $\s$.

\begin{result} \label{res2}
The exact star product in Result~\ref{res1} can be approximated as
\begin{eqnarray}  \label{pfunctstarprodapprox}
Q_A \star^{(-1)} Q_B
&= 
Q_A \exp[	(\overarrowethadj{\leftarrow} ) (\overarroweth{\rightarrow}) / (2J)] \, Q_B  + \mathcal{O}(J^{-3}) \\
P_A \star^{(1)} P_B
&= 
P_A \exp[-	(\overarroweth{\leftarrow} ) (\overarrowethadj{\rightarrow}) / (2J)]  \, P_B + \mathcal{O}(J^{-1}).
\end{eqnarray}
One obtains from Proposition~\ref{proposition1}
a more convenient approximation
\begin{eqnarray}  \label{pfunctstarprodapproxderivative}
Q_A \star^{(-1)} Q_B
&= 
Q_A \exp[ \,  \overleftarrow{\partial}_{\alpha} \overrightarrow{\partial}_{\alpha^*}] \, Q_B 
+ \mathcal{O}(J^{-1}) \\
P_A \star^{(1)} P_B
&= 
P_A \exp[ \,  - \overleftarrow{\partial}_{\alpha^*} \overrightarrow{\partial}_{\alpha}]  \, P_B + \mathcal{O}(J^{-1}),
\end{eqnarray}
which recovers the infinite-dimensional case in the large-spin limit if the functions $Q_A$, $Q_B$, $P_A$, and $P_B$
and their differentials remain non-singular in the limit.
\end{result}

\section{Transforming between phase-space representations \label{transformationssection}}
\subsection{Exact transformations \label{exacttransform}}
As detailed in \cite{koczor2017},
the convolution of the finite-dimensional distribution function $F_\rho(\Omega,s')$
with the spin-up state representation $F_{\spinvacuumpure}(\Omega,s)$
transforms between different representations and 
results in
a type-$(s{+}s'{-}1)$ distribution function (refer also to \eref{infinitedimensionalswitch})
\begin{equation}
\label{spinswitchdifferential}
F_\rho(\Omega,s{+}s'{-}1) = F_{\spinvacuumpure} (\theta,s) \ast F_\rho(\Omega,s')
=: \del{s} \, F_\rho(\Omega,s').
\end{equation}
We rely on \eref{spinswitchdifferential} to define the differential operator $\del{s}$
which will be often used in the following as a convenient notational shortcut for the convolution.
This differential operator satisfies the eigenvalue equation
$\del{s} \, \Y_{j m} = \gamma_{j}^{1{-}s} \, \Y_{j m}$
when applied to spherical harmonics \cite{koczor2017}.
\ref{Vandermonde} details how these eigenvalues  $\gamma_{j}^{1{-}s}$ can be written as a
$2J$-order polynomial in $j(j{+}1)$ which enables us to
specify $\del{s}$ as a polynomial in the differentials $\eth \ethadj$
using \eref{eigenfunct}:

\begin{result}
	\label{res3}
	The operator $\del{s}$ from \eref{spinswitchdifferential}
	can be specified as a $2J$-order polynomial 
	\begin{equation} \label{delequation}
	\del{s} F_\rho(\Omega,s') = \sum_{n=0}^{2J} c_n(s) \; (\eth \ethadj)^n F_\rho(\Omega,s')
	\end{equation}
	in terms of the differentials $\eth \ethadj$ (or equivalently $\ethadj \eth$),
	where the coefficients $c_n(s)$ are uniquely determined and can be
	computed analytically (refer to \ref{Vandermonde} for details).
\end{result}

The upper summation bound in \eref{delequation} can be enlarged to $4J$
if one performs a truncation of the higher-order spherical-harmonics terms in the resulting
phase-space distribution function, refer to \ref{Vandermonde}.

\subsection{Approximate transformations\label{sec:approx_transf}}

The eigenvalue equation for the transformation operator $\del{s}$ from \sref{exacttransform} 
is expanded into
\begin{equation}\label{asymptotic_expansion}
\del{s} \, \Y_{j m} = \, \sum_{n=0}^{\infty} [- \case{1{-}s}{ 4J} \, j(j{+}1) ]^n/n! \,  \Y_{j m}   + \mathcal{O}(J^{-1})
\end{equation}
by applying results of \ref{gammaconvergence}.
This enables the following approximations
which are also discussed in \fref{coeffapproxfig}(c):

\begin{result}
	\label{res4}
	Using the asymptotic expansion from \eref{asymptotic_expansion}, $\del{s}$ can be approximated by
	\begin{equation} \label{delapprox}
	\del{s} F_\rho(\Omega,s') = \exp[\, \case{1{-}s}{ 4J} \, \eth \ethadj] F_\rho(\Omega,s') + \mathcal{O}(J^{-1}).
	\end{equation}
	Proposition~\ref{proposition1} facilitates the approximation 
	\begin{equation} \label{delapproxderivative}
	\del{s} F_\rho(\Omega,s') 
	= \exp[\,  \case{1{-}s}{ 2} {\partial}_{\alpha^*}{\partial}_{\alpha} \, ] F_\rho(\Omega,s') + \mathcal{O}(J^{-1})
	\end{equation}
	in terms of the derivatives ${\partial}_{\alpha^*}$ and ${\partial}_{\alpha}$.
	This recovers the infinite-dimensional case from \eref{infinitedimensionalswitch} in the large-spin limit
	if $F_\rho(\Omega,s')$ and its differentials remain non-singular.
\end{result}

\section[Star products of s-parametrized phase spaces]{Star products of $s$-parametrized phase spaces \label{starproductssections}}
\subsection{The exact star product}

Generalizing from P and Q functions in \sref{pqfunctionstarprod}, the star product of general
$s$-parametrized phase spaces is now determined.
The differential operator $\del{s}$ can be used 
to translate the star product of
P or Q functions
to the star product of
arbitrary $s$-parametrized distribution functions. 
In particular, we apply Result~\ref{res1} to $Q_{AB} = Q_{A} \star^{(-1)} Q_{B}$
and  $P_{AB} = P_{A} \star^{(1)} P_{B}$ 
and use the substitutions 
\begin{eqnarray*}
\fl \qquad \qquad
& \del{s{+}2} Q_{AB} = F_{AB}(\Omega,s),
\;
Q_{A} = \del{{-}s} F_{A}(\Omega,s),
\;
Q_{B} = \del{{-}s} F_{B}(\Omega,s), \\
\fl \qquad \qquad
& \del{s} P_{AB} = F_{AB}(\Omega,s),
\;
P_{A} = \del{2{-}s} F_{A}(\Omega,s),
\;
P_{B} = \del{2{-}s} F_{B}(\Omega,s)
\end{eqnarray*}
from Result~\ref{res2} in order to compute the star product:
\begin{result}
	\label{res5}
The star product of two $s$-parametrized phase-space distribution
functions $F_{A}(\Omega,s)$ and $F_{B}(\Omega,s)$ is given by
either of the two equations
\begin{eqnarray}
\label{exactsparstarprod}
\fl \qquad
&  F_{A}(\Omega,s)
\star^{(s)} 
F_{B}(\Omega,s)
=
\del{s{+}2}
\{F_{A}(\Omega,s) \,
[\, \delover{\leftarrow}{{-}s} 
\; \star^{(-1)} 
\delover{\rightarrow}{{-}s}] \,
F_{B}(\Omega,s) \}\\ \label{exactsparstarprodP}
\fl \qquad
& F_{A}(\Omega,s)
\star^{(s)} 
F_{B}(\Omega,s)
=
\del{s}
\{ F_{A}(\Omega,s) \,
[\, \delover{\leftarrow}{2{-}s} 
\; \star^{(1)} 
\delover{\rightarrow}{2{-}s}] \,
F_{B}(\Omega,s) \}.
\end{eqnarray}
An explicit expansion can be calculated by expanding $\del{s}$ using \eref{delequation}
and $\star^{(\pm 1)} $ using \eref{result1exateq} and by applying the Leibniz identity
$\eth (fg) = (\eth f)g + f(\eth g)$. This results in an alternative form of
the exact star product in \eref{exactsparstarprod} and \eref{exactsparstarprodP}:
\begin{eqnarray} \label{orderingoperators}
f \star^{(s)}  g
=
\sum_{\underline{a},\, \underline{b}, \, \underline{c}, \, \underline{d}}
\lambda^{(s)}_{\underline{a}, \, \underline{b}, \, \underline{c}, \, \underline{d}}
[\dots (\ethadj)^{a_2} (\eth)^{b_1} (\ethadj)^{a_1} f ]
[\dots (\ethadj)^{d_2} (\eth)^{c_1} (\ethadj)^{d_1} g ].
\end{eqnarray}
The suitably chosen coefficients $\lambda^{(s)}_{\underline{a}, \, \underline{b}, \, \underline{c}, \, \underline{d}}$
are nonzero only if all of the indices $a_i, b_i, c_i, d_i$ are smaller than $2J{+}1$. Different values
for these coefficients are possible as the product of spin-weight raising and lowering
operators can be reordered using their commutators from \sref{spinweightedsphsection}.
But all possible values of the coefficients lead to the same unique result.
\end{result}
Although the choice of the coefficients in the finite sum in \eref{orderingoperators} is in general not unique 
due to the non-commutativity of $\eth$ and $\ethadj$, convenient formulas can be obtained
for explicit values of $J$ by reordering products of $\eth$ and $\ethadj$. The particular case of 
$J=1/2$ is discussed in \sref{spin12specialcase}. For large $J$, the star product can be
approximated using the commutative derivatives $\partial_{\alpha^*}$ and $\partial_{\alpha}$
from the infinite-dimensional case as discussed in \sref{starprodapprox}.
Also, \eref{exactsparstarprod} can always be used to
calculate the exact star product, but this approach consists of three consecutive steps,
as demonstrated in \sref{illustrativeexample}.

\subsection[The case of a single spin 1/2]{The case of a single spin $1/2$ \label{spin12specialcase}}
In the particular case of $J=1/2$, the \emph{exact} star product
in \eref{exactsparstarprod} can be simplified 
into a more convenient form by applying \eref{orderingoperators}.
Let $A$ and $B$ denote spin-1/2 operators
and their phase-space representations are given by
$f=F_{A}(\Omega,s)$ and $g=F_{B}(\Omega,s)$.
The star product is then determined by 
\begin{equation}
f \stars g
=
N_s \Proj  (f  \,
[ 1 +
a_s	(\overarrowethadj{\leftarrow}) (\overarroweth{\rightarrow} )
- b_s 	(\overarroweth{\leftarrow} ) (\overarrowethadj{\rightarrow}) 
]
\, g),
\end{equation}
where the $s$-dependent coefficients are $N_s = 2^{-\frac{s}{2}-\frac{1}{2}}$, 
\begin{equation*}
\fl \qquad \quad
a_s = \frac{1}{4} 3^{-s-\frac{1}{2}} [2\ 3^{s/2}-3^{s+\frac{1}{2}}+\sqrt{3} ],
\; \textup{ and } \;
b_s = \frac{1}{4} 3^{-s-\frac{1}{2}} [2\ 3^{s/2}+3^{s+\frac{1}{2}}-\sqrt{3} ].
\end{equation*}
The projection $\Proj:=1-\eth \ethadj/12 - \eth \ethadj \eth \ethadj/24$ 
removes superfluous terms in the spherical-harmonics decomposition,
i.e., contributions $\Y_{jm}$ with $j>1$ that do not correspond to
spin-$1/2$ distribution functions (refer to Result~2 in \cite{koczor2016}).
Note that for Wigner functions (i.e.\ $s=0$) the explicit form of the star product
can be calculated as (see, e.g., \cite{koczor2016}) 
\begin{equation}
W_A \, \star^{(0)} W_B 
=
\Proj  R\,   [ \sqrt{2 \pi} \, W_A W_B - \case{i}{2}\sqrt{\case{8 \pi}{3}} \, \{ W_A, W_B\}_S ]
\end{equation}
where $a_0 = b_0 = 1/(2 \sqrt{3})$
and $N_0 = 1/\sqrt{2}$.
For $J=1/2$, we have the radius $R=(4\pi)^{-1/2}$
and
the spherical Poisson bracket  has the form $i\{. , .\}_S = 
\overleftarrow{\partial}_\phi 
{(\sin\theta)}^{{-}1}
\overrightarrow{\partial}_\theta
-\overleftarrow{\partial}_\theta
(\sin\theta)^{{-}1}
\overrightarrow{\partial}_\phi$,
which should also be compared to \eref{poissonbracketdef}.

\subsection{Approximations of the star product \label{starprodapprox}}

Applying Result~\ref{res2} and Result~\ref{res4},
the exact star product in Result~\ref{res5} can be efficiently approximated
as detailed in the following:

\begin{result}
		\label{res6}
Let $f=F_{A}(\Omega,s)$ and $g=F_{B}(\Omega,s)$ denote the phase-space functions
of the spin-$J$ operators $A$ and $B$.
The star product in \eref{exactsparstarprod} can be expanded in terms
of spin-weight raising and lowering operators as
 (refer to \ref{proofofres3} for a proof)
\begin{equation}
\label{result3approspinweight}
f \star^{(s)} g
=
\sum_{n=0}^{4J}
\sum_{m=0}^{n}
\frac{c_{nm}(s)}{(2J)^n}
[\ethadjpower{m} \, \ethpower{n{-}m} f]
[\ethpower{m} \, \ethadjpower{n{-}m} g]
+ \mathcal{O}(J^{-1}),
\end{equation}
where the coefficients $c_{nm}(s)$ are defined in \ref{proofofres3}.
Similarly, the star product can be specified in terms of the derivatives
\begin{equation}
\label{generalstarprodapprox}
f \star^{(s)} g
=
f \exp [ \case{(1{-}s)}{2} \overleftarrow{\partial}_{\alpha} \overrightarrow{\partial}_{\alpha^*} 
-
\case{(1{+}s)}{2} \overleftarrow{\partial}_{\alpha^*} \overrightarrow{\partial}_{\alpha}]
\, g + \mathcal{O}(J^{-1}).
\end{equation}
The infinite-dimensional case from \Eref{infinitedimstarprod}
is recovered in the large-spin limit by applying Proposition~\ref{proposition1} if $f$ and $g$ and their
differentials remain non-singular.
\end{result}

Note that the first-order term (i.e.\ $n=1$) of the star product $\star^{(0)}$  in
\eref{result3approspinweight} is for the case of a Wigner function (i.e.\ $s=0$)
proportional to the spherical Poisson bracket \cite{koczor2016}
\begin{equation} \label{poissonbracketdef}
\fl \quad \{.,.\}_S:= i [
 (\overarrowethadj{\leftarrow}) (\overarroweth{\rightarrow} ) 
- (\overarroweth{\leftarrow} ) (\overarrowethadj{\rightarrow})
]/(2J)
=
\overleftarrow{\partial}_\phi 
{(2J\sin\theta)}^{{-}1}
\overrightarrow{\partial}_\theta
-\overleftarrow{\partial}_\theta
(2J\sin\theta)^{{-}1}
\overrightarrow{\partial}_\phi,
\end{equation}
which corresponds to the classical part of the time evolution \cite{koczor2016}.
The approximate star product can be also used to
derive efficient approximations of finite-dimensional phase-space
representations for large $J$ as illustrated in \sref{examplesection}.

\section{Time evolution of quantum states for a single spin \texorpdfstring{$J$}{J} \label{illustrativeexample}}
\subsection{Description of the time evolution using the star product\label{illustrativeexample_description}}

The time evolution of a quantum state $\rho$ for a single spin $J$
can be described in a phase space via the Moyal equation 
from \eref{moyaleq}. We discuss now the general structure
of the time evolution of phase-space functions along the lines of \cite{koczor2016}
and present an explicit example in \sref{illustrativeexamplesub}.
We refer to \cite{koczor2016} for further background and additional examples.
Substituting the $s$-parametrized
star product in \eref{moyaleq} with one of its forms from Result~\ref{res5}
yields the \emph{exact} equation of motion
for an arbitrary quantum state $F_\rho (\Omega,s)$ under a
Hamiltonian $F_{\mathcal{H}}(\Omega,s)$ as (refer, e.g.,
 \eref{exactsparstarprodP})
\begin{eqnarray}
\nonumber
\frac{\partial  F_\rho (\Omega,s)}{\partial t}
=
&- i \, \del{s}
\{  F_{\mathcal{H}} (\Omega,s) \,
[\, \delover{\leftarrow}{2{-}s} 
\; \star^{(1)} 
\delover{\rightarrow}{2{-}s}] \,
F_\rho (\Omega,s) \} \\
&+ i \, \del{s}
\{ F_\rho (\Omega,s) \,
[\, \delover{\leftarrow}{2{-}s} 
\; \star^{(1)} 
\delover{\rightarrow}{2{-}s}] \,
F_{\mathcal{H}} (\Omega,s) \}. \label{exact_time}
\end{eqnarray}
The use of this equation is illustrated in \sref{illustrativeexamplesub}
with the particular case of a Wigner function ($s=0$).
One can approximate this time evolution 
by substituting the $s$-parametrized
star product in \eref{moyaleq} with one of its approximations
from Result~\ref{res6}. This yields for $f:=F_{\mathcal{H}} (\Omega,s)$
and $g:= F_\rho (\Omega,s)$ an approximate equation
of motion up to an error of order $\mathcal{O}(J^{-1})$ as (e.g.)
\begin{equation} \label{approximatetimeexample}
\fl \;\;
\frac{\partial  F_\rho (\Omega,s)}{\partial t}
\approx \sum_{n=1}^{4J}
\sum_{m=0}^{n}
\frac{c_{nm}(s)}{(2J)^n}
 \{
- i
[\ethadjpower{m} \, \ethpower{n{-}m} f]
[\ethpower{m} \, \ethadjpower{n{-}m}g ]
+ i
[\ethadjpower{m} \, \ethpower{n{-}m} g]
[\ethpower{m} \, \ethadjpower{n{-}m} f ]
\}.
\end{equation}
Note that the first term of the summation ($n=1$)
coincides with the spherical Possion bracket from \eref{poissonbracketdef}
for the special case of Wigner functions ($s=0$) and corresponds to a semiclassical time evolution
\cite{koczor2016}. These semiclassical approximations have been widely used, refer to, e.g.,
\cite{starprod,klimov2005classical,zueco2007}.
Higher order contributions $n\geq 2$ are used to approximate quantum contributions
in the time evolution.

\subsection{Example of an explicit and exact time evolution for a single spin \texorpdfstring{$J$}{J}\label{illustrativeexamplesub}}
We discuss in this section an example 
for a single spin with arbitrary spin number $J$
which illustrates the application
of the exact star product from \eref{exactsparstarprod}.
Let us consider an experimental system with a
single spin $J$ that is controlled
as, e.g., in solid state nuclear magnetic resonance \cite{freude2000}.
The density operator of a single spin $J$ is in the thermal equilibrium 
given by $\rho_0 \propto \unity + \beta \mathcal{I}_z$
where $\beta$ depends on the temperature.
(We use the notations $\mathcal{I}_z = \T_{10}/(\sqrt{2}N_J)$
and $\mathcal{I}_+=\T_{11}/N_J$ with $N_J=1/\sqrt{2J(J{+}1)(2J{+}1)/3}$.)
We assume an effective Hamiltonian of the form $\mathcal{H}_{eff} := \omega (\mathcal{I}_+)^3 + \mathcal{H}_{res} $.
The first-order time evolution under this effective Hamiltonian is
given by the von-Neumann equation
$\frac{\partial \rho_0}{\partial t}
= -i \omega [ (\mathcal{I}_+)^3 , \mathcal{I}_z]
-i [ \mathcal{H}_{res} , \rho_0]  $,
and the commutator $[ (\mathcal{I}_+)^3 , \mathcal{I}_z]$  is proportional to $(\mathcal{I}_+)^3$.
The term $(\mathcal{I}_+)^3$ is responsible for creating multiple quantum coherences
which are often desirable. One can design experimental controls that 
maximize this contribution in the effective Hamiltonian \cite{freude2000}.

\begin{figure}
	\centering
	\includegraphics{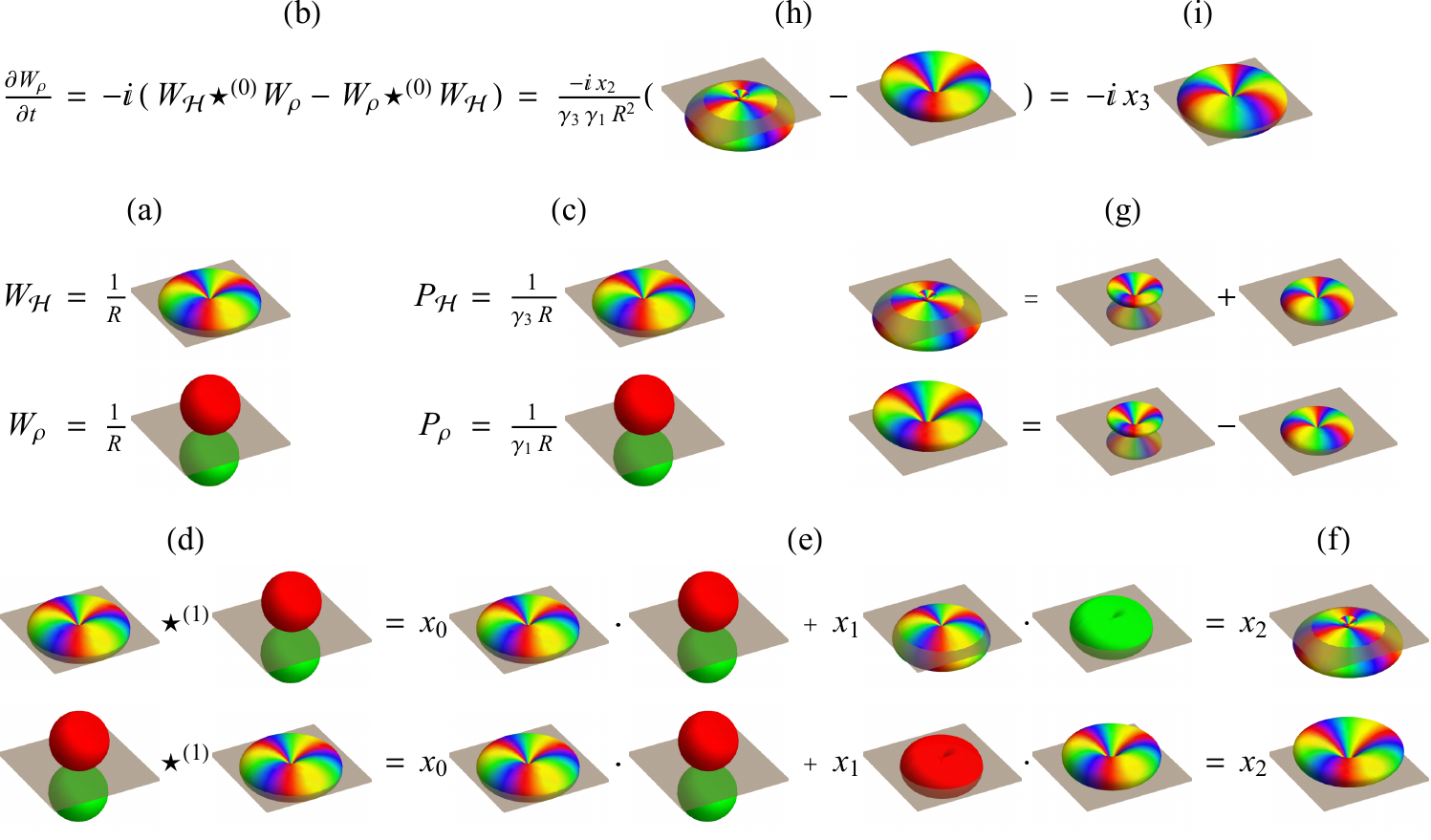}
	\caption{\label{illustration} Visualizing the calculation of the exact time evolution using  
	        the star product of Wigner functions
	        for a single spin with an arbitrary spin number J:
		(a) $W_{\mathcal{H}} :=  \case{1}{R} \Y_{3 3}(\theta,\phi)$
		and $W_{\rho} :=  \case{1}{R} \Y_{1 0}(\theta,\phi)$ and 
		(c) their corresponding P functions,
		(b) time evolution in the phase space via the Moyal equation \eref{starcomm},
		(d)-(f) star products $P_{\mathcal{H}}  \star^{(1)}   P_{\rho}$ and $  P_{\rho} \star^{(1)}  P_{\mathcal{H}} $
		using \eref{illustrationstrprod1}-\eref{illustrationstrprod2},
		(g) the resulting star products decompose into two spherical harmonics as in \eref{illustrationdecompose};
		(h) substituting (f) back into the Moyal equation (b) leads to 
		(i) where the symmetric contribution from (g) is canceled out.
		The prefactors $x_0,x_1,x_2,x_3$ are used to rescale some of the plots.
	}
\end{figure}

Equivalently, the time evolution of the density operator $ \rho := \T_{1 0}$
under the Hamiltonian $\mathcal{H} := \T_{3 3}$
can be calculated for a single spin with arbitrary spin number $J$
directly in a phase-space representation.
The corresponding spin Wigner functions 
$W_{\mathcal{H}} :=  \case{1}{R} \Y_{3 3}(\theta,\phi)$
and
$W_{\rho} :=  \case{1}{R} \Y_{1 0}(\theta,\phi)$
are specified in terms of spherical harmonics,
see \fref{illustration}(a).
The time evolution of the Wigner function $W_{\rho}$ 
is described in the phase space by the Moyal equation (cf.\ \cite{koczor2016})
\begin{equation}
\label{starcomm}
\frac{\partial W_{\rho}}{\partial t}
=
-i \, W_{\mathcal{H}} \, \star^{(0)} W_{\rho}
+i \, W_{\rho} \, \star^{(0)} W_{\mathcal{H}}
\end{equation}
by  relying on the star commutator 
which is determined in terms of star products,
see \fref{illustration}(b).
The star product $W_{\mathcal{H}} \, \star^{(0)} W_{\rho}$ is evaluated via Result~\ref{res5}.
Using \eref{exactsparstarprodP}, we get
\begin{equation}
\label{illustration1}
 W_{\mathcal{H}} \, \star^{(0)} W_{\rho}
=
\del{0}
( \, W_{\mathcal{H}} \,
[\, \delover{\leftarrow}{2} 
\; \star^{(1)} 
\delover{\rightarrow}{2}] \,
 W_{\rho} \,
),
\end{equation}
where $\del{2}$ is determined by Result~\ref{res3} (refer to \Eref{spinswitchdifferential})
and transforms the Wigner functions to the corresponding P
functions $P_{\mathcal{H}} = W_{\mathcal{H}} \delover{\leftarrow}{2} =\case{1}{R \gamma_{3}} \Y_{3 3}(\theta,\phi) $
and $P_{\rho} =\, \delover{\rightarrow}{2} W_{\rho} =\case{1}{R \gamma_{1}} \Y_{1 0}(\theta,\phi) $,
see \fref{illustration}(c). Note
that spherical harmonics $\Y_{jm}$ are eigenfunctions of $\del{2}$
with eigenvalues $\gamma^{-1}_{j}$.
The right hand side of \eref{illustration1} is then equal to
$\del{0}
( \, P_{\mathcal{H}} \star^{(1)}   P_{\rho} \, )$, for
which the star product of P functions is computed using
\eref{result1exateq} in Result~\ref{res1}.
This yields (see \fref{illustration}(d-f))
\begin{eqnarray}
\label{illustrationstrprod1}
P_{\mathcal{H}}  \star^{(1)}   P_{\rho}
&=
\case{\Y_{3 3} \star^{(1)} \Y_{1 0}}{R^2 \gamma_{1} \gamma_{3}}
=
\case{1}{R^2 \gamma_{1} \gamma_{3}}
[ \lambda^{(1)}_0 \Y_{3 3}  \Y_{1 0}
+
\lambda^{(1)}_1 (\eth \Y_{3 3})  (\ethadj \Y_{1 0}) ],\\
\label{illustrationstrprod2}
P_{\mathcal{H}}  \star^{(1)}   P_{\rho}
&=
\case{1}{R^2 \gamma_{1} \gamma_{3}}
[ \lambda^{(1)}_0 \Y_{3 3}  \Y_{1 0}
-
\lambda^{(1)}_1 2\sqrt{6}
\,
 \Y^1_{3 3} \Y^{-1}_{1 0} ].
\end{eqnarray}
The coefficients 
$ \lambda^{(1)}_0$ and  $\lambda^{(1)}_1$
are defined in \eref{exactstarprodceffs}
and the operators $\eth$ and $\ethadj$
from \eref{ETHoperatorC} are responsible for
raising and lowering the spin weight of the
spherical harmonics.

Products of spin-weighted spherical
harmonics can be decomposed into sums 
$\Y_{3 3}  \Y_{1 0}
=
\case{1}{2 \sqrt{3 \pi}}
\Y_{4 3}$
and
$\Y^{ 1}_{3 3} \Y^{- 1}_{1 0}
=
- \case{3}{4 \sqrt{2\pi}}  \Y_{3 3} 
+ \case{1}{4 \sqrt{2\pi}}  \Y_{4 3}$ 
of spherical
harmonics.
The corresponding star product of Wigner functions is
obtained by rescaling spherical harmonics $\Y_{j m}$ by $\gamma_{j}$,
which results in (refer to \fref{illustration}(g-h))
\begin{eqnarray}
\label{illustrationdecompose}
W_{\mathcal{H}}  \star^{(0)}   W_{\rho}
&= \case{1}{R^2 \gamma_{1} \gamma_{3}} 
[+ \case{\lambda^{(1)}_1 3 \sqrt{3}}{2 \sqrt{\pi}} \, \gamma_{3}   \Y_{3 3} 
+  (\case{\lambda^{(1)}_0}{2 \sqrt{3 \pi}} {-}\case{\lambda^{(1)}_1 \sqrt{3}}{2 \sqrt{\pi}} ) \, \gamma_{4} \Y_{4 3} ] \\
W_{\rho}   \star^{(0)}   W_{\mathcal{H}} 
&=
\case{1}{R^2 \gamma_{1} \gamma_{3}}
[ - \case{\lambda^{(1)}_1 3 \sqrt{3}}{2 \sqrt{\pi}}  \, \gamma_{3} \Y_{3 3} 
+  (\case{\lambda^{(1)}_0}{2 \sqrt{3 \pi}} {-}\case{\lambda^{(1)}_1 \sqrt{3}}{2 \sqrt{\pi}} ) \, \gamma_{4} \Y_{4 3} ].
\end{eqnarray}
Note that $\gamma_{4}=0$ for $J < 2$ which is responsible
for truncating the spherical-harmonics decomposition \cite{koczor2016}, 
refer to \ref{Vandermonde}.
The final result determining the time evolution
is obtained via the star commutator
from \eref{starcomm}, and we obtain (for arbitrary J)
\begin{equation}
\frac{\partial W_{\rho}}{\partial t}
= - i \, \case{3 \, \sqrt{3}}{\sqrt{\pi}} \case{\lambda^{(1)}_1}{ R^2 \gamma_{1}}  \Y_{3 3}.
\end{equation}

\subsection{Extending the example to an arbitrary quantum state\label{illustrativeexample_arb}}

Applying the same Hamiltonian $W_{\mathcal{H}}  = \case{1}{R} \Y_{3 3}$ to an arbitrary quantum state $g:= W_\rho$,
we could apply \eref{exact_time} to determine the time evolution exactly. But we will consider here only
the approximate time evolution (see \eref{approximatetimeexample})
\begin{equation*} 
\fl \quad
\frac{\partial  W_{\rho}}{\partial t}
\approx   \case{1}{R}  \sum_{n=1}^{4J}
\sum_{m=0}^{n}
\frac{c_{nm}(0)}{(2J)^n}
\{
- i
[\ethadjpower{m} \, \ethpower{n{-}m} \Y_{3 3}]
[\ethpower{m} \, \ethadjpower{n{-}m}g ]
+ i
[\ethadjpower{m} \, \ethpower{n{-}m} g]
[\ethpower{m} \, \ethadjpower{n{-}m} \Y_{3 3} ]
\},
\end{equation*}
where one can apply \eref{SPHdefininition} to simplify
differentials. For example, one obtains (up to a factor)
the spin-weighted spherical harmonics 
\begin{equation}
\ethadjpower{m} \, \ethpower{n{-}m} \Y_{3 3}
\propto
\cases{
     \Y^{n-2m}_{3 3}
	&for $|n-m| \leq 3$ and $|n-2m| \leq 3$,
	\\
	0
	& otherwise.
	\\}
\end{equation}
This highlights that most of the terms in the sum vanish for a general spin
number $J$. But this example makes it also apparent that a
semiclassical approximation \cite{starprod,klimov2005classical,zueco2007}
that restricts the summation to $n=1$ will neglect relevant quantum contributions.

\section{One example of photon-added coherent states \label{examplesection}}

Creation and annihilation operators are widely used 
and account for numerous non-classical effects including, for example,
photon-added coherent states which were demonstrated experimentally
\cite{zavatta2004,zavatta2005,zavatta2007,barbieri2010,kumar2013,agarwal1991}.
Photon-added coherent states are obtained from coherent states 
$| \alpha_0 \rangle := \mathcal{D}(\alpha_0) | 0 \rangle$
as $a^\dagger | \alpha_0 \rangle$ by applying
the creation operator $a^\dagger$.
The inversely translated version of these quantum states
is created from the vacuum state by applying the operator $Q(\alpha_0)$ 
and one has
\begin{equation}
\label{infidimexample}
 \pht =  Q(\alpha_0) | 0 \rangle
\;
\textup{ with }
\;
Q(\alpha_0):= \case{1}{\sqrt{1+ |\alpha_0|^2}} \mathcal{D}(\alpha_0)^{-1}  a^\dagger \mathcal{D}(\alpha_0).
\end{equation}
Phase-space representations of these photon-added coherent states can be obtained
using the star products of the individual phase-space representations 
\begin{equation}
F_{\pht}(\alpha,s) =  F_{Q(\alpha_0)} \stars F_{| 0 \rangle}(\alpha,s) \stars (F_{Q(\alpha_0)})^*,
\end{equation}
where the Gaussian function $F_{| 0 \rangle}(\alpha,s) = 2\exp{[-2 \alpha \alpha^*/(1 {-} s)]}/(1{-}s)$
represents the vacuum state \cite{cahill1969}
and  $F_{Q(\alpha_0)} :=(\alpha^* {+}\alpha_0^*)/\sqrt{1{+} |\alpha_0|^2}$ corresponds to the creation operator.
Applying the star product from \eref{infinitedimstarprod} yields
\begin{equation}
\fl \quad
F_{\pht}(\alpha,s) =[\alpha {+}\alpha_0 {-} \case{1{+}s}{2} \partial_{\alpha^*}]
[\alpha^* {+}\alpha_0^* {-} \case{1{+}s}{2} \partial_{\alpha}]
F_{| 0 \rangle}(\alpha,s)
=:
\overline{\mathcal{Q}}(\alpha_0)  \mathcal{Q}(\alpha_0) F_{| 0 \rangle}(\alpha,s),
\end{equation}
where the second equality describes the photon creation in the shifted phase space
in terms of the differential operators 
$\mathcal{Q}(\alpha_0) \overline{\mathcal{Q}}(\alpha_0)= \overline{\mathcal{Q}}(\alpha_0) \mathcal{Q} (\alpha_0)$.
Setting $\alpha_0=0$ yields the phase-space equivalent of the creation operator, i.e.,
essentially Bopp operators \cite{bopp1956}
\begin{equation}
\mathcal{Q} := \mathcal{Q}(0) = [\alpha^* - \case{1{+}s}{2} \partial_{\alpha}]
\;
\textup{ and }
\;
\overline{\mathcal{Q}} := \overline{\mathcal{Q}}(0)=  [\alpha - \case{1{+}s}{2} \partial_{\alpha^*}].
\end{equation}
For the number state $| n \rangle$, one can 
calculate the  phase-space functions
\begin{equation}
\fl \;
F_{| n \rangle}(\alpha,s) =
\case{1}{n!}
[\overline{\mathcal{Q}}  \mathcal{Q}  ]^n
 F_{| 0 \rangle}(\alpha,s)
 \;
 \textup{ and }
 \;
 F_{| n_1 \rangle \langle n_2|}(\alpha,s) =
 \case{1}{\sqrt{n_1!} \sqrt{n_2!}}
 (\overline{\mathcal{Q}})^{n_2}
 (\mathcal{Q} )^{n_1}
 F_{| 0 \rangle}(\alpha,s)
\end{equation}
corresponds to tilted projectors $| n_1 \rangle \langle n_2|$ which span a complete,
orthonormal basis \cite{cahill1969}.

\begin{figure}
	\centering
	\includegraphics{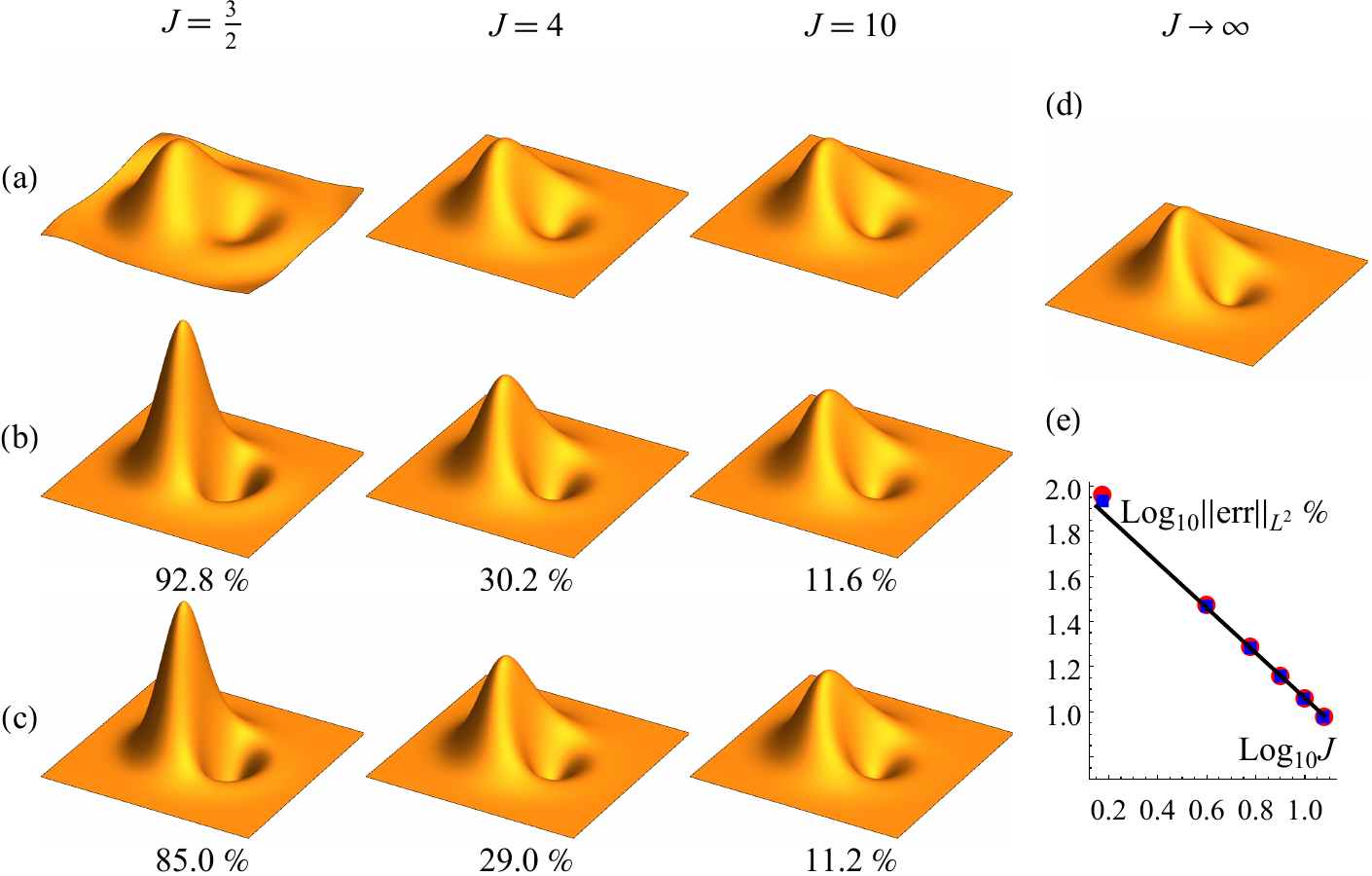}
	\caption{\label{exmplefig} (a) Plots of the exact phase-space representations $F_{\spinraised}(\Omega,0)$ of
		$\spinraised$ for increasing spin numbers $J$ using \eref{exampledefinition}.
		(b)-(c) Approximations of  $F_{\spinraised}(\Omega,0)$
		using \eref{exampleresult} and  the Gaussian function $F_{\vacuum}(\Omega,0)$, also
		applying the \emph{approximate} star product in \eref{result3approspinweight} for (b)
		and \eref{generalstarprodapprox} for (c).
		Both  (b) and (c) converge to (a) in order $\mathcal{O}(J^{-1})$ as they
		approach their infinite-dimensional counterpart (d) from \eref{infidimexample}.
		The approximation errors in the $L^2$ norm shown below the plots are
		graphed in (e) for $J \in \{3/2, 4, 6, 8, 10, 12 \}$, where red dots and blue squares refer,
		respectively, to (b) and (c) and the black line shows the expected scaling $\mathcal{O}(J^{-1})$
		with a suitable prefactor.
		The distance from the origin in these planar plots
		represent the arc distance from the north pole in the spherical phase space.
		The rotation parameter in \eref{exampledefinition} has been chosen as $\Omega_0 = (\theta_0,\phi_0) = (1.2 /\sqrt{J/2}$,0)
		and the translation parameter in \eref{infidimexample} is given by $\alpha_0 = 1.2$.
		Plot limits are $\pm 2$. Refer also to \Fref{convergencefig} and \Fref{convergenceballfig}.
	}
\end{figure}

We define finite-dimensional analogues of photon-added coherent states
in the form
\begin{equation}
\label{exampledefinition}
\spinraised:= K(\Omega_0) | JJ \rangle
\;
\textup{ with }
\;
K(\Omega_0) := \case{N}{\sqrt{2J}} \mathcal{R}^{-1}(\Omega_0) \mathcal{J}_- \mathcal{R}(\Omega_0),
\end{equation}
where $| JJ \rangle$ is the spin-up state
and $\mathcal{J}_-/\sqrt{2J}$ is the finite-dimensional analogue of the creation operator
which approaches $a^\dagger$ in the large-spin limit. Refer to Table~1 in \cite{arecchi1972atomic}
and  \ref{exampleappendix} for details.
Following the infinite-dimensional characterization,
the phase-space representation $F_{\spinraised}$
can be written in terms of the exact star product 
of the individual phase-space representations as (see Result~\ref{res5})
\begin{equation}
\label{exmaplestarprod}
F_{\spinraised}(\Omega,s) =  \FF \stars F_{\vacuum}(\Omega,s) \stars (\FF)^*,
\end{equation}
where the Gaussian-like function $F_{\vacuum}(\Omega,s)$ represents the
spin-up state and $\FF$ is the
phase-space representation of the rotated lowering operator $K(\Omega_0)$.
Refer to \Fref{exmplefig}(a) for plots of the (unapproximated) $F_{\spinraised}(\Omega,s)$ for $s=0$.
Using, e.g., the exact star product in  \eref{exactsparstarprod},
the phase-space representation 
\begin{figure}
	\centering
	\includegraphics{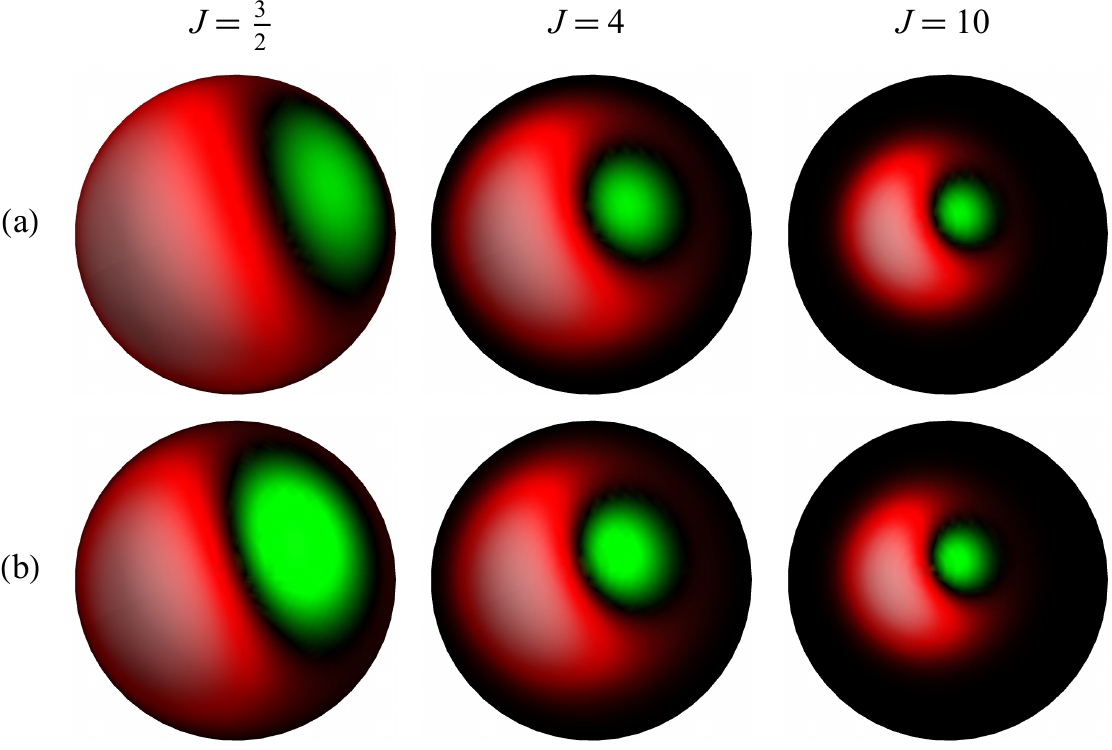}
	\caption{\label{convergenceballfig}
		Alternative representation of \fref{exmplefig}(a)-(b).
		(a) Plots of the exact phase-space representations $F_{\spinraised}(\Omega,0)$ of
		$\spinraised$ for increasing $J$ using \eref{exampledefinition}.
		(b) Approximations of $F_{\spinraised}(\Omega,0)$
		via  \eref{exampleresult} and \eref{result3approspinweight} assuming a Gaussian $F_{\vacuum}(\Omega,0)$.
		(b) converges to (a) in order $\mathcal{O}(J^{-1})$.
		Spherical functions $f(\theta,\phi)$ are plotted on  
		the sphere where the brightness represents the absolute value
		and the colors red (dark gray) and
		green (light gray) depict positive and negative values, respectively.
	}
\end{figure}
\begin{equation}
\label{exampleresult}
F_{\spinraised}(\Omega,s)
=
 \lp   \lpc \,
F_{\vacuum}(\Omega,s)
=  \lpc  \lp \,
F_{\vacuum}(\Omega,s)
\end{equation}
of the excited coherent state
$F_{\spinraised}(\Omega,s)$ is obtained by applying the 
differential operators $\lp$ and  $\lpc$ 
which are defined via their action on phase space functions $f$, i.e.,
\begin{equation}
\lp f :=  \FF \stars f
\; \textup{ and } \;
\lpc f :=  f \stars (\FF)^*.
\end{equation}
Approximations of these operators can be used to approximate 
$F_{\spinraised}(\Omega,s)$ by applying them to a Gaussian approximation of
$F_{\vacuum}(\Omega,s)$ as given in \cite{koczor2017}.
Approximations of  $\lp$ and  $\lpc$ are calculated
in terms of spin-weighted spherical harmonics and their raising and lowering operators
in \ref{exampleappendix} using the \emph{approximate} star products
in \eref{result3approspinweight} and \eref{generalstarprodapprox}.
These approximations converge 
in \fref{exmplefig}
to the exact phase-space
functions,
and the infinite-dimensional
case from \eref{infidimexample} is recovered in the large-spin limit.

Comparable to the infinite-dimensional case, the unrotated ladder operator
$\lpn:=\mathcal{K}(0)$ is specified in \ref{exampleappendix} in 
terms of spin-weighted spherical
harmonics and their raising and lowering operators.
One can calculate representations
\begin{equation}
\label{Dickestate}
F_{|Jm\rangle}(\Omega,s)
\propto
[ \lpn   \lpcn ]^{J-m} \,
F_{\vacuum}(\Omega,s)
\end{equation}
of the Dicke state $|Jm\rangle$
using the operators $\lpn$ and $\lpcn$,
and similar representations 
\begin{equation}
\label{projectors}
F_{|Jm_1\rangle \langle Jm_2 |}(\Omega,s)
\propto
 (\lpcn)^{J-m_2} ( \lpn  )^{J-m_1}  \,
F_{\vacuum}(\Omega,s)
\end{equation}
are obtained for the tilted projectors $|J m_1\rangle \langle Jm_2 |$.
All of these states have typically quite complicated spherical-harmonics
expansions which are challenging to calculate for large values of $J$.
Approximations based on the star product in \eref{generalstarprodapprox}
facilitate efficient calculations of these and similar  phase-space representations for large $J$.

We want to close this section by remarking that general 
(infinite-dimensional) $s$-parametrized phase spaces
naturally appear in experimental homodyne measurements
\cite{Leonhardt97,zavatta2004,zavatta2005,zavatta2007,barbieri2010,kumar2013} of 
the discussed photon-added coherent states. As is explained in \cite{Leonhardt93},
the relevant experiment yields $s$-parametrized phase-space functions
with $s=-(1{-}\xi)/\xi$ for a detector efficiency of $\xi$. This provides
an example for the occurrence of $s$-parametrized phase spaces beyond the
particular cases of $s\in\{-1,0,1\}$.

\section{Generalization to coupled spins \label{generalization}}

The explicit form of the star product for Wigner functions of coupled spin-$1/2$ systems was detailed
in Result~3 of \cite{koczor2016}. Building on results in \cite{koczor2016},
we outline how to generalize these results to $s$-parameterized phase spaces of
coupled spins $J$. We consider two operators $A$ and $B$
in a system of $N$ coupled spins $J$.
Their phase-space representations are determined similarly as in Result~1 of \cite{koczor2016} and
can be calculated as
\begin{equation}
F_A(s,\theta_1,\phi_1, \ldots, \theta_N,  \phi_N):=
 \tr [  A \bigotimes_{k=1}^{N} \, \mathcal{R}(\theta_k,  \phi_k)  M_s  \mathcal{R}^{\dagger}(\theta_k,  \phi_k) ],
\end{equation}
where the transformation kernel in Result~1 of \cite{koczor2016} is expressed here in terms of
rotated parity operators from \eref{PSrepDefinition}.
We generalize the star product described in Result~\ref{res5} to the star product
\begin{equation}
\label{exactcoupledstar}
	F_A \star F_B :=    F_A  ( \prod_{k=1}^{N}   [\stars]^{ \{k \} } ) F_B
\end{equation}
of phase-space representations for
coupled spins by applying Result~3 of \cite{koczor2016},
where $\stars$ is the star product from Result~\ref{res5} and $[\stars]^{ \{k \} }$ describes
that the star product acts only on the variables $\theta_k$ and  $\phi_k$.
\Eref{exactcoupledstar} completely specifies the exact star product for a system
of $N$ interacting spins $J$, and the corresponding approximations via
Result~\ref{res6} can be conveniently expressed using
the commutativity of partial derivatives.
For example, the approximate star product in terms of
the derivatives $\partial_{\alpha}$ and $\partial_{\alpha^*}$
from \eref{generalstarprodapprox} is generalized for coupled spins to
\begin{equation}
\label{coupledstarprodapprox}
\fl \qquad F_A  \star^{(s)} F_B
=
F_A  \exp [ \sum_{k=1}^{N} (\case{(1{-}s)}{2} \overleftarrow{\partial}_{\alpha_k} \overrightarrow{\partial}_{\alpha_k^*} 
-
\case{(1{+}s)}{2} \overleftarrow{\partial}_{\alpha_k^*} \overrightarrow{\partial}_{\alpha_k}
)]
\, F_B + \mathcal{O}(J^{-1}).
\end{equation}
The equation of motion for Wigner functions, i.e., the Moyal equation from
\eref{starcomm}, can be consequently established for a system of coupled spins $J$
using \eref{coupledstarprodapprox} as
\begin{equation*}
\fl \qquad
i \frac{\partial W_{\rho}}{\partial t}
=
 \, W_{\mathcal{H}} \, \star^{(0)} W_{\rho}
- \, W_{\rho} \, \star^{(0)} W_{\mathcal{H}}
=
W_{\mathcal{H}} [
 e^{i \{.,.\}/2}
- e^{-i \{.,.\}/2}
]W_{\rho}
+ \mathcal{O}(J^{-1}),
\end{equation*}
where $\{.,.\}:=\sum_{k=1}^N g_k$ and
$g_k:=-i \overleftarrow{\partial}_{\alpha_k} \overrightarrow{\partial}_{\alpha_k^*} 
+
 i \overleftarrow{\partial}_{\alpha_k^*} \overrightarrow{\partial}_{\alpha_k}$ specify
a Poisson bracket acting on the variables $\alpha_k$ and $\alpha_k^*$.
This results in the expansion
\begin{eqnarray}
i \frac{\partial W_{\rho}}{\partial t}
&=
W_{\mathcal{H}}[
2 \sum_{n=0, \atop \textrm{\tiny $n$ odd}}
(-i \{.,.\}/2)^{n}/n!
] W_{\rho} + \mathcal{O}(J^{-1}) \\
&=
W_{\mathcal{H}}[
-i \sum_{k=1}^N g_k
+ \case{i}{24}
\hspace{-3mm}
\sum_{k_1,k_2,k_3=1}^N 
\hspace{-3mm}
g_{k_1}g_{k_2}g_{k_3}
+ \cdots] W_{\rho}
+ \mathcal{O}(J^{-1}).
\end{eqnarray}
Using Proposition~\ref{proposition1}, the differential operators $-g_{k}$
can be replaced by the spherical Poisson brackets
$p_{k}:=\{.,.\}_S^{ \{k \}  }$
from \eref{poissonbracketdef},
which results in the time evolution
\begin{equation}
\fl \quad
i \frac{\partial W_{\rho}}{\partial t} =
W_{\mathcal{H}}[
\, i  \, \sum_{k=1}^N p_k
- \case{i}{24} \hspace{-6mm}
\sum_{k_1,k_2,k_3=1 \atop {k_\mu \neq k_\nu \textrm{ \tiny for $\mu \neq \nu$}} }^N
\hspace{-6mm}
	 p_{k_1}p_{k_2}p_{k_3}  + \cdots
{-} \case{i}{24} \sum_{k=1}^N p_{k}^3
+ \cdots] W_{\rho}
+\mathcal{O}(J^{-1}).
\end{equation}
The first two terms (before the first dots) can be directly compared to the ones appearing in the 
star product of coupled spins $1/2$ in Result~4 of \cite{koczor2016}.
The leading term corresponds to the classical equation of motion,
and the following terms in the expansion are ordered according to their degree of non-locality
as proposed in Result~4 of \cite{koczor2016}.

\section{Conclusion \label{conclusion}}
We have derived the exact star product for continuous 
$s$-parametrized phase-space
representations of single spins $J$ in terms of spin-weighted spherical
harmonics and their raising and lowering operators. Our construction
naturally recovers the well-known case of infinite-dimensional quantum systems
in the limit of large spin numbers $J$. Based on approximations of spin-weighted
spherical harmonics, we have derived convenient formulas for approximating
star products which, beyond the time evolution, can be useful for efficiently
calculating phase-space representations for large spin numbers. We have 
illustrated our methods and their application in concrete examples.
We have finally outlined how the presented formalism can be extended to coupled spin systems.
In summary, we have established a complete phase-space description for 
finite-dimensional quantum systems and their time evolution.

\ack
B.K.\ acknowledges financial support from the scholarship program of the 
Bavarian Academic Center for Central, Eastern and Southeastern Europe (BAYHOST).
R.Z. and S.J.G. acknowledge support from the Deutsche
Forschungsgemeinschaft (DFG) through Grant No.\ Gl 203/7-2.

\appendix

\section{Expansions of products of tensor operators \label{productexpansionappendix}}

Adopting the notation of  \cite{koczor2016}, the product of two irreducible 
tensor operators can be---similarly as in \eref{topdecomposition}---expanded as  (see \cite{varshalovich})
\begin{equation} \label{ProdTop}
\T_{j_1 m_1} \T_{j_2 m_2}
= 
\sum_{L=|j_1-j_2|}^{n} 
{}^JQ_{j_1 j_2 L} \,
C^{LM}_{j_1 m_1 j_2 m_2}
\T_{L M}.
\end{equation}
The  upper limit $n:=\min( j_1{+}j_2, 2J )$ of the summation 
is bounded by $2J$. We have set $M:=m_1{+}m_2$ and also use
the Clebsch-Gordan coefficients $C^{LM}_{j_1 m_1,j_2 m_2}$ \cite{messiah1962}.
The coefficients $^{J}Q_{j_1 j_2 L}$ from \cite{koczor2016}
are proportional to Wigner $6$-$j$ symbols \cite{messiah1962} and 
depend only on $j_1$, $j_2$, and $L$, but are independent of $m_1$, $m_2$, and $M$.

Similarly, the product of any two spin-weighted spherical harmonics can be decomposed into
a sum of spin-weighted spherical harmonics (see Eq.~2.54 in \cite{Del2012})
\begin{eqnarray} \label{ProdOfTwoSpinSPH}
\Y^{\s_1}_{j_1 m_1} \Y^{\s_2}_{j_2 m_2}  \nonumber
= & \sum_{L=|j_1-j_2|}^{j_1+j_2}  (-1)^{M+\s_3}  \sqrt{\case{(2j_1{+}1)(2j_2{+}1)(2L{+}1)}{4\pi}}  \\
&  \times \myThreeJ{j_1 & j_2 & L}{m_1 & m_2 & -M}
\myThreeJ{j_1 & j_2 & L}{-\s_1 & -\s_2 & \s_3}
\Y^{\s_3}_{L M}
\end{eqnarray}
where the Wigner $3$-$j$ symbols \cite{messiah1962} are used. The values of $M=m_1+m_2$ and $\s_3=\s_1+\s_2$ 
are bounded by
$-L \leq M \leq L$ and $-L \leq \s_3 \leq L$.
Substituting the left-hand side of this equation with the definition of spin-weighted spherical harmonics
$\Y^{\s_1}_{j_1 m_1}$ and $\Y^{\s_2}_{j_2 m_2}$ from \eref{SPHdefininition} while also assuming that $\s_1=-\s_2=:\s$,
one obtains the relation
\begin{eqnarray} \label{ethstarprodexpansion}
( \ethpower{\s} \, \Y_{j_1 m_1} ) ( \ethadjpower{\s} \, \Y_{j_2 m_2})  \nonumber
=  &\sum_{L=|j_1-j_2|}^{j_1+j_2}   (-1)^{M+\s}  \sqrt{\case{(2j_1{+}1)(2j_2{+}1)(2L{+}1)}{4\pi}} 
\\
&  \times x_{\s \, j_1}^{ j_2}
\myThreeJ{j_1 & j_2 & L}{m_1 & m_2 & -M}
\myThreeJ{j_1 & j_2 & L}{-\s & \s & 0}
\Y_{L M},
\label{SpinWeightedSphericalHarmonicsProduct}
\end{eqnarray}
where the factor $x_{\s \, j_1}^{ j_2}$ can be obtained from \eref{SPHdefininition} and is determined by
\begin{equation}
x_{\s \, j_1}^{ j_2}= \sqrt{\case{(j_2{+}\s)!(j_1{+}\s)!}{(j_2{-}\s)!(j_1{-}\s)!}}.
\end{equation}
Finally, the explicit form of the factor $\kappa$ in 
\eref{spinweighteddecompositionA}-\eref{spinweighteddecompositionB} is now given by (with $M=m_1{+}m_2$)
\begin{eqnarray} \label{kappaexplicit}
{}^{\s}\kappa_{j_1m_1,j_2m_2}^{L}  \nonumber
=  &  (-1)^{M+\s}  \sqrt{\case{(2j_1{+}1)(2j_2{+}1)(2L{+}1)}{4\pi}} 
\\
&  \times x_{|\s| \, j_1}^{ j_2}
\myThreeJ{j_1 & j_2 & L}{m_1 & m_2 & -M}
\myThreeJ{j_1 & j_2 & L}{-\s & \s & 0}
\Y_{L M}.
\end{eqnarray}

\section{Proof of Result~\ref{res1} \label{proofofresult1}}
We prove now Result~\ref{res1}.
Both formulas in \eref{result1exateq} must satisfy
the defining property \eref{starproductcondition} of the star product.
The expansions from \eref{spinweighteddecompositionA}-\eref{spinweighteddecompositionB}
result in the condition 
\begin{equation*}
K_{jm,j'm'}^{L} = [\frac{\gamma_L}{\gamma_j \gamma_{j'}}]^{\pm 1} \frac{1}{R}
\sum_{\s=0}^{2J} \lambda^{( \pm 1)}_\s
\, \, {}^{\pm \s}\kappa_{j_1m_1,j_2m_2}^{L}
\end{equation*}
for $\lambda^{( \pm 1)}_\s$ which holds 
for every $j,m,j',m'$ with $j,j'\leq 2J$.
This specifies an overdetermined linear system of equations for $\lambda^{( \pm 1)}_\s$
which can be recognized as the matrix-vector equation
$K = \kappa^{( \pm 1)} \,
\lambda^{( \pm 1)} $. Here, the vector $\lambda^{( \pm 1)} $ has the entries 
$\lambda^{( \pm 1)}_\s$ and every entry $K_i$ of the vector $K$
is given by a value of $K_{j_i m_i ,j'_i m'_i}^{L_i}$ with $i \in \{1, 2, \ldots, (2J{+}1)^5 \}$.
The corresponding matrix $\kappa^{( \pm 1)}$ has the dimension
$(2J{+}1) \times (2J{+}1)^5$ and rank $(2J{+}1)$. This linear system has a unique,
exact solution and one obtains the coefficients in \eref{exactstarprodceffs}.

\section{Asymptotic expansion of weight factors \label{gammaconvergence}}

Detailed expansion formulas for \sref{sec:approx_star} and \sref{sec:approx_transf}  are computed in the following.
The coefficients in \eref{exactstarprodceffs} can be  expanded 
into the form
\begin{equation*}
\fl \qquad
\lambda^{(-1)}_\eta \eta! = \frac{(2J{-}\eta)! }{ (2J)!} = \prod_{k=0}^{\s-1} (2J{-}k)^{-1} = \prod_{k=0}^{\s-1} [(2J)^{-1} + k (2J)^{-2} + \mathcal{O}((2J)^{-3})],
\end{equation*}
where the second equality follows from the Taylor expansion $(a+b)^{-1} = 1/a - b/a^2 +b^2/a^3 + \cdots$
with $a:=2J$ and $b:=-k$ and $|b|<a$.
Collecting the error terms as $(2J)^{-\eta+1} \; \sum_{k=0}^{\s-1} [k \; (2J)^{-2}]=(2J)^{-\eta-1}[\s(\s{-}1)]/2$ 
yields the formula
\begin{equation*}
\lambda^{(-1)}_\s \s!   = (2J)^{-\s}  + (2J)^{-\eta-1}[\s(\s{-}1)] + \mathcal{O}((2J)^{-\s-2}).
\end{equation*}
This results in the asymptotic expansion
in \eref{lambdaapprox0}-\eref{lambdaapprox1}.
Similarly, $\lambda^{(1)}_\eta $ is expanded as
\begin{equation*}
\fl \;
\lambda^{(1)}_\eta (-1)^\s \s!  = \frac{ 2J\,  (2J)! }{ (2J{+}\s{+}1)!}
=
2J \prod_{k=1}^{\s+1} (2J{+}k)^{-1} 
=
2J \prod_{k=1}^{\s+1} [(2J)^{-1} + k (2J)^{-2} + \mathcal{O}((2J)^{-3})],
\end{equation*}
which simplifies to the asymptotic expansion (which is used in \eref{lambdaapprox2})
\begin{equation*}
\lambda^{(1)}_\s (-1)^\s \s!   = (2J)^{-\s}  + (2J)^{-\eta-1} \case{(\s+1)(\s+2)}{2} + \mathcal{O}((2J)^{-\s-2}).
\end{equation*}

The coefficients in the definition of spin-weighted spherical harmonics in \eref{SPHdefininition}
can similarly be expanded as
\begin{equation*}
\fl \qquad \qquad
\sqrt{{(j{-}\s)!}/{(j{+}\s)!}}
=
\prod_{k=-\s+1}^{\s} (j+k)^{1/2} 
= \prod_{k=-\s+1}^{\s} [ j^{1/2}  + \case{k}{2 j^{1/2} }+ \mathcal{O}(j^{-3/2})]
\end{equation*}
where the second equality is obtained from the Taylor expansion
$(a+b)^{1/2} = a^{1/2} + b/(2 a^{1/2})  -  b^2/(8 a^{3/2}) + \cdots$
with $a:=j$ and $b:=k$ and $b<a$. This yields the expansion	
\begin{equation}
\label{spinweightedSPHprefactorexpansion}
\sqrt{{(j{-}\s)!}/{(j{+}\s)!}}
=
j^{\s}  + \case{\s}{2 j^{1/2} } + \mathcal{O}(j^{-5/2})	.
\end{equation}

Following similar arguments, the factor $\gamma_j$, which is defined in \eref{gammafactor},
can be in terms of $j(j{+}1)$ expanded into the exponential function 
\begin{equation}\label{gammaapprox}
\fl \qquad 
\gamma_j = \exp{[ - j(j{+}1)/(4J) ]} + \mathcal{O}(J^{-1})
= \sum_n [- j(j{+}1)/(4J)]^n/n! + \mathcal{O}(J^{-1}).
\end{equation}
This expansion is used in \eref{asymptotic_expansion} to derive an approximation of the
operator $\del{s}$.

\section{Proof of Result~\ref{res3} and the associated expansion coefficients \label{Vandermonde}}
We now prove Result~\ref{res3} and determine the corresponding expansion coefficients $c_n(s)$.
The coefficients $c_n(s)$ are uniquely determined by the values of $\gamma_{j}^{1{-}s}$ and the condition
\begin{equation}
\label{transformationoplineareq}
\gamma_{j}^{1{-}s} =  \sum_{n=0}^{2J} c_n(s) \; [-j(j{+}1)]^n
\;
\textup{ for }
\;
0 \leq j \leq 2J.
\end{equation}
This yields the linear system 
$V \, c(s) 
=
\gamma(s)
$
of equations
where $V$ is the Vandermonde matrix
with entries $[V]_{nj}:=[-j(j{+}1)]^n$ and its inverse $V^{-1}$ 
can be computed analytically \cite{bjorck1970}.
The entries of the vectors $c(s)$ and $\gamma(s)$
are given by $c_n(s)$ and $\gamma_{j}^{1{-}s}$, respectively.
The exact, unique solution is determined by
$V^{-1} \, \gamma(s)
= 
c(s)
$.

Simultaneously truncating the spherical-harmonics decomposition can also
be achieved by enlarging the summation upper limit in
\eref{transformationoplineareq} to $4J$. In that case, one has
$\del{s} \Y_{jm} = 0$ for $2J < j \leq 4J$.
Alternatively, a projection operator $\mathcal{P}_J$ from Result~2
of \cite{koczor2016} can be applied to spin-$J$ phase-space
representations, where
$\mathcal{P}_J := \sum_{n=0}^{4J} p_n \; (\eth \ethadj)^n$
and the coefficients $p_n$ are computed from the linear system of equations
\begin{equation}
\sum_{n=0}^{4J} p_n \; [-j(j{+}1)]^n
=
\cases{
1
&for $0 \leq j \leq 2J$
\\
0
&for $2J < j \leq 4J$
}
\end{equation}
which is determined by the inverse Vandermonde matrix $V^{-1}$.

\section{Asymptotic expansion of differential operators \label{diffopconvergence}}
\subsection{Expansion formulas using polar and arc-length parametrizations}
In this section, we show how the operators $\eth$ and $\ethadj$
approach their infinite-dimensional counterparts given by the
derivatives $\partial_{\alpha^*}$ and $\partial_\alpha$.
We consider the polar parametrization 
$\alpha = r e^{i\phi}$
of the complex plane
with
$r=\sqrt{\alpha^* \alpha}$ and $\phi:=\arg{\alpha}$.
Using
${\partial}_{\alpha} = \case{\partial r}{\partial \alpha} \partial_{r}
+  \case{\partial \phi}{\partial \alpha}\partial_{\phi}$,
the derivatives
$\partial_\alpha$ and $\partial_{\alpha^*}$ can be expressed in the polar parametrization
by substituting $\case{\partial r}{\partial \alpha} = \case{1}{2} e^{-i\phi}$
and $\case{\partial \phi}{\partial \alpha} =\case{- i}{2} e^{-i\phi} /r$
which results in
\begin{eqnarray}
\label{alphaderivativechainrule}
{\partial}_{\alpha} = 
e^{-i\phi} \case{1}{2} [ \partial_{r} -  i /r \; \partial_{\phi} ]
\;
\textup{ and }
\;
{\partial}_{\alpha^*} = 
e^{i\phi} \case{1}{2} [ \partial_{r} +  i /r \; \partial_{\phi} ].
\end{eqnarray}
Applying these formulas one obtains formulas for powers of derivatives:
\begin{eqnarray}
\label{prodeth}
\fl \quad
[{\partial}_{\alpha^*}]^\s = e^{i \s \phi} \prod_{k=0}^{\s-1} \case{1}{2} [ -k/r + \partial_{r} +  i /r \; \partial_{\phi} ],
\;
[{\partial}_{\alpha}]^\s = e^{-i \s \phi} \prod_{k=0}^{\s-1} \case{1}{2}  [ -k/r + \partial_{r} -  i /r \; \partial_{\phi} ].
\end{eqnarray}

For comparison,
we apply the spin-weight raising and lowering differential operators from
\eref{ETHoperatorA} and \eref{ETHoperatorB}
and obtain
\begin{eqnarray} \label{productexpansion1}
\fl \qquad \qquad
[\eth / \sqrt{2J} ]^\s \, \Y_{j m}   &= 
(-1)^\s \prod_{k=0}^{\s-1} \case{1}{2}
 ( - k \case{\cos{\theta}}{\sqrt{J/2} \sin{\theta}} +  \case{\partial_\theta}{ \sqrt{J/2} }
+ \case{i}{ \sqrt{J/2} \sin{\theta } } \; \partial_\phi ) 
 \Y_{j m}    \\ \label{productexpansion2}
\fl \qquad \qquad
[ \ethadj / \sqrt{2J} ]^\s \, \Y_{j m}   &= 
(-1)^\s \prod_{k=0}^{\s-1} \case{1}{2}
 ( - k \case{ \cos{\theta}}{\sqrt{J/2} \sin{\theta}} +  \case{\partial_\theta}{ \sqrt{J/2} }
 - \case{i}{\sqrt{J/2} \sin{\theta}} \; \partial_\phi )
\Y_{j m}  .
\end{eqnarray}
The arc-length parametrization $\alpha=\sqrt{J/2} \, \theta \, e^{{-}i\phi}$
(see, e.g., \cite{koczor2017}) implies that $r=\sqrt{\alpha^* \alpha} = \sqrt{J/2} \, \theta$.
The following terms from \eref{productexpansion1} and \eref{productexpansion2}
can be expanded by applying their Taylor series and substituting $\theta=r / \sqrt{J/2}$:
\begin{equation}
\fl \;
\label{sinuslimit}
 \case{i}{ \sqrt{J/2} \sin{(r / \sqrt{J/2})} } = \case{i}{r} + \case{i r}{3 J} + \mathcal{O}( J^{-3/2})
\;
\textup{ and }
\;
\case{\cos{(r / \sqrt{J/2})}}{\sqrt{J/2} \sin{(r / \sqrt{J/2})}}   = \case{1}{r} - \case{2 r}{3 J} + \mathcal{O}( J^{-3/2}).
\end{equation}
Substituting these expansions back into \eref{productexpansion1} and \ref{productexpansion2} results in
\begin{eqnarray*}
\fl \qquad [\eth / \sqrt{2J} ]^\s \, \Y_{j m}   & = 
(-1)^\s \prod_{k=0}^{\s-1} \case{1}{2}
[ - k (\case{1}{r}{ }- \case{2 r}{3 J})  +  \partial_r
+ (\case{i}{r} {+} \case{i r}{3 J})  \partial_\phi + \mathcal{O}( J^{-3/2})   
]
\Y_{j m} ,   \\ 
\fl \qquad [ \ethadj / \sqrt{2J} ]^\s \, \Y_{j m}  & = 
(-1)^\s \prod_{k=0}^{\s-1} \case{1}{2}
[ - k (\case{1}{r} {-} \case{2 r}{3 J}) +  \partial_r
- (\case{i}{r} {+} \case{i r}{3 J}) \; \partial_\phi  + \mathcal{O}( J^{-3/2}) 
]
\Y_{j m}.
\end{eqnarray*}
Note that $\partial_\phi \Y^{\s}_{j m} = im \Y^{\s}_{j m}$.
The expressions in the parentheses can 
up to an error term $\epsilon := \case{r}{3J}(2 k {-} m)$
be transformed
into terms that are directly comparable to \eref{prodeth}:
\begin{eqnarray} \label{diffopwitherrorsA}
\fl \qquad \quad
[\eth / \sqrt{2J} ]^\s \, \Y_{j m}   & = 
(-1)^\s \prod_{k=0}^{\s-1} \case{1}{2}
[ - k /r  +  \partial_r
+ i/r  \partial_\phi + \epsilon + \mathcal{O}( J^{-3/2})   
]
\Y_{j m}  ,  \\ \label{diffopwitherrorsB}
\fl \qquad \quad
[ \ethadj / \sqrt{2J} ]^\s \, \Y_{j m}  & = 
(-1)^\s \prod_{k=0}^{\s-1} \case{1}{2}
[ - k /r +  \partial_r
- i/r  \; \partial_\phi - \epsilon  + \mathcal{O}( J^{-3/2}) 
]
\Y_{j m}.
\end{eqnarray}
We compare \eref{prodeth} with \eref{diffopwitherrorsA} and \eref{diffopwitherrorsB},
apply $\sum_{k=0}^{\s-1} \epsilon/2 = \case{r}{6J}(1{+}\s)(\s{-}m)$, and denote
the residual error terms by $\zeta:=\mathcal{O}( J^{-3/2} [\eth / \sqrt{2J} ]^{\s-1} )$
and $\bar{\zeta}:=\mathcal{O}( J^{-3/2} [\ethadj / \sqrt{2J} ]^{\s-1} )$. This leads to
\begin{eqnarray}\label{derivativeapproxA}
\fl \qquad \qquad
& [\eth / \sqrt{2J} ]^\s \, \Y_{j m}   = [ (-1)^\s e^{- i \s \phi} ({\partial}_{\alpha^*})^\s
+
\case{r(1+\s)(\s-m)}{6J}   [\eth / \sqrt{2J} ]^{\s-1}  + \zeta ] \, \Y_{j m} 
\\ \label{derivativeapproxB}
\fl \qquad \qquad
&[ \ethadj / \sqrt{2J} ]^\s \, \Y_{j m}   = 
[(-1)^\s e^{ i \s \phi} ({\partial}_{\alpha})^\s
-
\case{r(1+\s)(\s-m)}{6J} [ \ethadj / \sqrt{2J} ]^{\s-1}  + \bar{\zeta} ] \, \Y_{j m}.
\end{eqnarray}
Substituting this expansion into the definition of spin-weighted spherical harmonics in
\eref{SPHdefininition}, one obtains for a fixed arc length $\alpha$ the forms
(refer to \eref{SPHapproxA}-\eref{SPHapproxB} and \Fref{approxfig}(a-b))
\begin{eqnarray} 
&\Y^\s_{j m}  
-
(-1)^\s e^{ - i \s \phi} ({\partial}_{\alpha^*})^\s \, \Y_{j m}
  \propto |\alpha| (  j \sqrt{ J})^{-1}  \, \Y^{\s-1}_{j m} 
  +  \mathcal{O}( J^{-1}) \; \textup{ and} \\
& \Y^{-\s}_{j m}  
- 
e^{ i \s \phi} ({\partial}_{\alpha})^\s \, \Y_{j m}
\propto
  |\alpha|  ( j \sqrt{ J})^{-1} \, \Y^{-\s+1}_{j m}
    +  \mathcal{O}( J^{-1}).
\end{eqnarray}
The difference on the left-hand side vanishes in the limit of infinite $J$
for every bounded $\alpha$, as
the spin-weighted spherical harmonics $\Y^{-\s\pm 1}_{j m}$ on the right-hand
side are bounded, i.e., $|\Y^{-\s \pm 1}_{j m}(\theta,\phi)| < \infty$.

We now describe a general criterion 
(see \sref{ethapproximations})
for spherical functions and their
differentials to be bounded.
Assume now that the spherical function $f= f(\theta,\phi)=\sum f_{jm} \Y_{j m}(\theta,\phi)$
is bounded, i.e., $|f(\theta,\phi)| < \infty$.
Note that the expansion coefficients might depend on $J$.
Also, assume that the differentials are bounded, i.e.,
 $ |\eth^\s f(\theta,\phi)| < \infty$ and $ |\ethadj^\s f(\theta,\phi)| < \infty$,
which translates to 
\begin{eqnarray*}
| ( \eth / \sqrt{2J} )^\s  f(\theta,\phi)| = |\sum \sqrt{{(2J)^\s (j{-}\s)!}/{(j{+}\s)!}}^{-1} f_{jm} \Y^\s_{j m}(\theta,\phi)| < \infty\\
| ( \ethadj / \sqrt{2J} )^\s  f(\theta,\phi)| = |\sum \sqrt{{(2J)^\s (j{-}\s)!}/{(j{+}\s)!}}^{-1} f_{jm} \Y^{-\s}_{j m}(\theta,\phi)| < \infty.
\end{eqnarray*}
We emphasize that $f$ and all of its derivatives are bounded
if there are only a finite number of non-zero expansion coefficients $f_{jm}$
or if the expansion coefficients $|f_{jm}|$ decay faster in $j$ than the coefficients
$\sqrt{{(j{-}\s)!}/{(j{+}\s)!}} \approx j^{-\s}$ from \eref{spinweightedSPHprefactorexpansion}.
Applying \eref{diffopwitherrorsA} and \eref{diffopwitherrorsB} to the spherical function $f$
one gets for a fixed arc length $|\alpha|$ that
\begin{eqnarray*} 
&[ (\eth / \sqrt{2J} )^\s - (-1)^\s e^{- i \s \phi} ({\partial}_{\alpha^*})^\s] \, f(\alpha)
\propto
|\alpha| J^{-1} \,  [\eth/\sqrt{2J}]^{\s-1} f(\alpha) 	
\\
& [ ( \ethadj / \sqrt{2J} )^\s - (-1)^\s e^{ i \s \phi} ({\partial}_{\alpha})^\s   ] \, f(\alpha)
\propto
|\alpha|  J^{-1} \,  [\ethadj/\sqrt{2J}]^{\s-1} f(\alpha)  .
\end{eqnarray*}
This difference clearly vanishes if  $g=|\alpha|  \, [\eth/\sqrt{2J}]^{\s-1} f(\alpha)$
remains bounded in the limit of infinite $J$.
(The assumption could be weakened such that the growth of
the absolute value of $g$ in $J$ is slower than $\mathcal{O}(J)$.)
For a fixed $\alpha$,
this expansion of the action of spin-weight lowering and raising operators
has the convergence rate $\mathcal{O}(J^{-1})$,
refer to Proposition~\ref{proposition1}.

We now describe when 
spherical functions and their differentials are in general
bounded in the $L^2$ norm, and this information is
utilized in \sref{ethapproximations}.
The asymptotic behavior of the difference function
can be measured in the $L^2$ norm.
Assume that the square-integrable spherical function $f=f(\theta,\phi)=\sum f_{jm} \Y_{j m}(\theta,\phi)$
observes $R^2 \sum |f_{jm}|^2 = 1$.
Also asssume that its differentials are square integrable. 
Applying the orthonormality of spin-weighted spherical harmonics, this translates 
to 
\begin{eqnarray} 
\fl \qquad \qquad || ( \eth / \sqrt{2J} )^\s f(\theta,\phi)||_{L^2} &= R^2\sum ((2J)^\s {(j{-}\s)!}/{(j{+}\s)!})^{-1} |f_{jm}|^2  < \infty, \\
\fl \qquad \qquad || ( \ethadj / \sqrt{2J} )^\s f(\theta,\phi)||_{L^2} &=R^2 \sum ((2J)^\s {(j{-}\s)!}/{(j{+}\s)!})^{-1} |f_{jm}|^2  < \infty.
\end{eqnarray}
Note that $f$ and all of its derivatives are square integrable for finite $J$
if there are only a finite number of non-zero expansion coefficients $f_{jm}$
or if the expansion coefficients $|f_{jm}|^2$ decay faster in $j$ than
${(j{-}\s)!}/{(j{+}\s)!} \approx j^{-2\s}$ from \eref{spinweightedSPHprefactorexpansion}.
Applying \eref{diffopwitherrorsA} and \eref{diffopwitherrorsB}, the norm of the difference
is given by
\begin{eqnarray*} 
\fl & \qquad  || \, [ (\eth / \sqrt{2J} )^\s - (-1)^\s e^{- i \s \phi} ({\partial}_{\alpha^*})^\s] \, f(\theta,\phi)||_{L^2}
\propto
 J^{-1} \,  || \, |\alpha|  \, [\eth/\sqrt{2J}]^{\s-1} f(\theta,\phi)||_{L^2}
\\
\fl & \qquad || \,  [ ( \ethadj / \sqrt{2J} )^\s - (-1)^\s e^{ i \s \phi} ({\partial}_{\alpha})^\s   ] \, f(\theta,\phi)||_{L^2}
\propto
 J^{-1} \,  || \, |\alpha| \, [\ethadj/\sqrt{2J}]^{\s-1} f(\theta,\phi)||_{L^2}.
\end{eqnarray*}
This difference clearly vanishes if the norm $|| \, |\alpha|  \, [\eth/\sqrt{2J}]^{\s-1}	 f(\theta,\phi)||_{L^2}$
remains bounded in the large-spin limit.
(This assumption can be weakened such that the growth of this norm in $J$ is slower than $\mathcal{O}(J)$.)
Refer to Proposition~\ref{proposition1}.

Alternatively, the following expansions can be derived from \eref{diffopwitherrorsA} and \eref{diffopwitherrorsB}:
\begin{eqnarray*}
& 
[ (\eth / \sqrt{2J} )^\s - (-1)^\s e^{- i \s \phi} ({\partial}_{\alpha^*})^\s] f(\alpha) 
\propto
J^{-1} |\alpha|   \frac{\partial^{\s-1} f(\alpha) }{({\partial}_{\alpha^*})^{\s-1}}  , \\
& [ ( \ethadj / \sqrt{2J} )^\s - (-1)^\s e^{ i \s \phi} ({\partial}_{\alpha})^\s   ] \, f(\alpha)
\propto
 J^{-1}  |\alpha|  \frac{\partial^{\s-1} f(\alpha) }{({\partial}_{\alpha})^{\s-1}} .
\end{eqnarray*}
These differences vanish in the limit of infinite $J$ if the derivatives remain bounded, i.e.,
\begin{equation}
| \,  |\alpha|   \frac{\partial^{\s-1} f(\alpha,\alpha^*) }{({\partial}_{\alpha^*})^{\s-1}}| < \infty
\; \textup{ and } \;
| \,  |\alpha|   \frac{\partial^{\s-1} f(\alpha,\alpha^*) }{({\partial}_{\alpha^*})^{\s-1}}| < \infty.
\end{equation}
In addition, the $L^2$ norm of the differences vanishes if the derivatives remain square integrable, i.e.,
\begin{equation}
|| \,  |\alpha|   \frac{\partial^{\s-1} f(\alpha,\alpha^*) }{({\partial}_{\alpha^*})^{\s-1}}||_{L^2} < \infty
\; \textup{ and } \;
|| \,  |\alpha|   \frac{\partial^{\s-1} f(\alpha,\alpha^*) }{({\partial}_{\alpha^*})^{\s-1}}||_{L^2} < \infty,
\end{equation}
or if the growth of the norm and the absolute value in $J$ is slower than $\mathcal{O}(J)$.
This is used in Proposition~\ref{proposition1}.

Following similar arguments, asymptotic expansions for products of differentials
from Proposition~\ref{proposition2} are obtained in the formulas
\begin{eqnarray}
|f   [(\overarrowethpower{\leftarrow} ) 
(\overarrowethadjpower{\rightarrow})/(2J)^\s 
]
g
-
f
(\overleftarrow{\partial}_{\alpha^*})^\s (\overrightarrow{\partial}_{\alpha})^\s 
g|
\propto  J^{-1} \; \textup{ and}
\\
|| \, f [  (\overarrowethpower{\leftarrow} ) 
(\overarrowethadjpower{\rightarrow}) /(2J)^\s
-
(\overleftarrow{\partial}_{\alpha^*})^\s (\overrightarrow{\partial}_{\alpha})^\s ]
 \, g
||_{L^2}	
\propto
J^{-1},
\end{eqnarray}
and the two formulas are also valid for the conjugate derivatives.
Consider two square-integrable functions
$f$ and $g$ with the additional constraint that the product $f g$
as well as the products of differentials $|\alpha| (\ethpower{\s} f) (\ethadjpower{\s} g)$ are square integrable.
The $L^2$-norm convergence then holds (refer to Proposition~\ref{proposition2}).

\subsection[The product of the spin-weight raising and lowering operators]{The product $\eth \ethadj$ of the spin-weight raising and lowering operators}

We now derive the second part of Proposition~\ref{proposition1}.
The derivatives ${\partial}_{\alpha^*}{\partial}_{\alpha}$ in the
polar parametrization are expanded into 
\begin{equation}
{\partial}_{\alpha^*}{\partial}_{\alpha}
=
[- 1/r +  \partial_{r} +  i /r \; \partial_{\phi} ]
[ \partial_{r} -  i /r \; \partial_{\phi} ]/4.
\end{equation}
Similarly, the expansion of the operator $\eth \ethadj /(2J)$ is given by
\begin{equation}
\fl \quad
\eth \ethadj /(2J) \, \Y_{j m}   
=
[- \case{\cos{\theta}}{\sqrt{2J} \sin{\theta}}  + \partial_\theta/\sqrt{2J}
+ \case{i}{\sqrt{2J} \sin{\theta} } \; \partial_\phi ]
[ \, \partial_\theta/\sqrt{2J} - \case{i}{\sqrt{2J} \sin{\theta}} \; \partial_\phi ] 
\, \Y_{j m}  .
\end{equation}
Applying the expansions from \eref{sinuslimit}
and the parametrization $\theta=r / \sqrt{J/2}$
yields
\begin{eqnarray*}
\eth \ethadj /(2J) \, \Y_{j m}   
= &
[- \case{1}{r} - \case{2 r}{3 J}  + \partial_r
+  (\case{i}{r} {+} \case{i r}{3 J} )  \partial_\phi + \mathcal{O}( J^{-3/2})  ] \\
& \times
[ \partial_r - (\case{i}{r} {+} \case{i r}{3 J} ) \partial_\phi + \mathcal{O}( J^{-3/2}) ] 
\, \Y_{j m} /4 .
\end{eqnarray*}
Now separating the terms and expanding the action $\partial_\phi \Y_{j m}$ results in
\begin{eqnarray*}
\eth \ethadj /(2J) \, \Y_{j m}   
= &
[- \case{1}{r}  + \partial_r
+  \case{i}{r}   \partial_\phi - \case{(m+2) r}{3 J}+ \mathcal{O}( J^{-3/2})  ]
\\  & \times
[ \, \partial_r - \case{i}{r}\; \partial_\phi - \case{m r}{3 J}  + \mathcal{O}( J^{-3/2}) ] 
\, \Y_{j m} /4 .
\end{eqnarray*}
Finally, we obtain for a bounded spherical function $f$ that
\begin{equation*}
\fl \qquad \qquad [ \eth \ethadj /(2J)
-
{\partial}_{\alpha^*}{\partial}_{\alpha}
] f
\propto  |\alpha| J^{-1}
\; \textup{ and } \;
|| [ \eth \ethadj /(2J) 
-
{\partial}_{\alpha^*}{\partial}_{\alpha}
]\,f  ||_{L^2}
\propto
 |\alpha| J^{-1} .
\end{equation*}
The norm or the absolute value vanish in the large-spin limit if both the
function $f$ and its differentials are bounded
or square integrable in the limit, see Proposition~\ref{proposition1}.

\section{Proof of Result~\ref{res6} \label{proofofres3}} 
We prove Result~\ref{res6}.
Substituting the approximations of $\del{s}$ from \eref{delapprox}
and of $\star^{(-1)} $ form \eref{pfunctstarprodapproxderivative} into
\eref{exactsparstarprod} yields the formula
\begin{eqnarray*}
& f \,
[\, \delover{\leftarrow}{{-}s} 
\; \star^{(-1)} 
\delover{\rightarrow}{{-}s}] \,
g \\
&=
f \, \exp[\,  \case{1{+}s}{ 2} {\overleftarrow{\partial}}_{\alpha^*}{\overleftarrow{\partial}}_{\alpha} \, ] 
 \exp[ \,  \overleftarrow{\partial}_{\alpha} \overrightarrow{\partial}_{\alpha^*}] 
 \exp[\,  \case{1{+}s}{ 2} {\overrightarrow{\partial}}_{\alpha^*}{\overrightarrow{\partial}}_{\alpha} \, ]\,g
  + \mathcal{O}(J^{-1})  \\
&=
f \, \exp[\,   \case{1{+}s}{ 2} {\overleftarrow{\partial}}_{\alpha^*}{\overleftarrow{\partial}}_{\alpha}
+  \overleftarrow{\partial}_{\alpha} \overrightarrow{\partial}_{\alpha^*}
+ \case{1{+}s}{ 2} {\overrightarrow{\partial}}_{\alpha^*}{\overrightarrow{\partial}}_{\alpha} \, ]\,g
+ \mathcal{O}(J^{-1}),
\end{eqnarray*}
where the second equality follows from the commutativity of partial derivatives. Using the Leibniz rule
of partial derivatives 
\begin{equation}
{\partial}_{\alpha^*}{\partial}_{\alpha} (fg)
=
({\partial}_{\alpha^*}{\partial}_{\alpha} f)g + f( {\partial}_{\alpha^*}{\partial}_{\alpha} g)
+({\partial}_{\alpha^*} f )({\partial}_{\alpha} g)  + ({\partial}_{\alpha} f) ({\partial}_{\alpha^*} g )
\end{equation}
results in
a convenient description for the action of the approximation of $\del{s} fg$:
\begin{equation}
\fl \quad
 \exp[\, - \case{1{+}s}{ 2} {\partial}_{\alpha^*}{\partial}_{\alpha} \, ] fg
 =
f  \exp[\,  - \case{1{+}s}{ 2} \left(
{\overleftarrow{\partial}}_{\alpha^*}{\overleftarrow{\partial}}_{\alpha} 
{+}  {\overrightarrow{\partial}}_{\alpha^*}{\overrightarrow{\partial}}_{\alpha}
{+} \overleftarrow{\partial}_{\alpha^*} \overrightarrow{\partial}_{\alpha}
{+}  \overleftarrow{\partial}_{\alpha} \overrightarrow{\partial}_{\alpha^*}
 \right) \, ] g.
\end{equation}
Substituting this into $\del{s{+}2}\left( f \,
	[\, \delover{\leftarrow}{{-}s} 
	\; \star^{(-1)} 
	\delover{\rightarrow}{{-}s}] \,
	g \right) $,
one obtains \eref{generalstarprodapprox}
which is expanded as
\begin{equation}
f \star^{(s)} g
=
\sum_{n=0}^{\infty}
\sum_{m=0}^{n}
c_{nm}(s) \, 
( \partial_{\alpha}^m \, {\partial}_{\alpha^*}^{n-m} f)
(\partial_{\alpha^*}^m   \, \partial_{\alpha}^{n-m} g)
 + \mathcal{O}(J^{-1}),
\end{equation}
where $c_{nm}(s)$ are the expansion coefficients of the
exponential $ \exp [ \case{(1{-}s)}{2} a - \case{(1{+}s)}{2} b]
=\sum_{n=0}^{\infty} \sum_{m=0}^{n} c_{nm}(s) \, a^m b^{n-m}$ for commutative $a$ and $b$.
Using the polar parametrization from \eref{prodeth},
the derivatives $ \partial_{\alpha}^n \, {\partial}_{\alpha^*}^m$
can be represented as
\begin{eqnarray}
\fl
\qquad 
\partial_{\alpha}^n \, {\partial}_{\alpha^*}^m= e^{i(m- n) \phi} \prod_{\eta=-m}^{n-m-1} \case{1}{2}  [ -\s/r + \partial_{r} -  i /r \; \partial_{\phi} ]
\prod_{\eta=0}^{m-1} \case{1}{2} [ -\s/r + \partial_{r} +  i /r \; \partial_{\phi} ].
\end{eqnarray}
One applies arguments from \ref{diffopconvergence} and the 
expansion 
\begin{equation}
( \partial_{\alpha}^n \, {\partial}_{\alpha^*}^m f)
(\partial_{\alpha^*}^n   \, \partial_{\alpha}^m g)
=
(2J)^{-n-m} (\ethadjpower{n} \ethpower{m} f	) (\ethpower{n} \ethadjpower{m}  g)
 + \mathcal{O}(J^{-1})
\end{equation}
of the differential operators can be established
which finally yields \eref{result3approspinweight}.

\section{Details for the example in \sref{examplesection} \label{exampleappendix}}

We discuss some details for the example in \sref{examplesection}.
The normalization factor $N$ in \eref{exampledefinition}
can be computed using
$1/N^2 = \langle JJ | K^\dagger  K |JJ \rangle $
where
\begin{eqnarray}
K &=\mathcal{R}^{-1}(\Omega_0) \mathcal{J}_- \mathcal{R}(\Omega_0)/\sqrt{2J} 
\nonumber \\
&=[
\mathcal{J}_- D^1_{-1,-1}(\Omega') 
+  \mathcal{J}_z D^1_{0,-1}(\Omega') / \sqrt{2}
+ \mathcal{J}_+ D^1_{1,-1}(\Omega')]/\sqrt{2J}.\label{eq:contr}
\end{eqnarray}
Here, $D^j_{m,m'}$ are Wigner D-matrix elements \cite{cohen1991quantum}.
All the contributions in \eref{eq:contr} vanish
except for $\langle JJ | \mathcal{J}_+ \mathcal{J}_- |JJ \rangle = 2J$
and $\langle JJ | \mathcal{J}_z \mathcal{J}_z |JJ \rangle = J^2$.
Finally, one obtains $1/N^2= \cos(\theta/2)^2 [1 + 2 J - (2 J {-} 1) \cos(\theta)] / 2$.

The phase-space representation $\FF$ of the operator $K$ from \eref{exampledefinition}
can be specified in terms of spherical harmonics as \cite{koczor2016}
\begin{eqnarray*}
	& \FF
	 =
	c_s \mathcal{R}(\Omega_0) \Y_{1,-1}(\Omega) \\
	& =
	c_s [
	\Y_{1,-1}(\Omega) D^1_{-1,-1}(\Omega_0) 
	+ \Y_{1,0}(\Omega) D^1_{0,-1}(\Omega_0)
	+ \Y_{1,1}(\Omega) D^1_{1,-1}(\Omega_0)],
\end{eqnarray*}
where the rotation can be written in terms of Wigner D-matrices
and the prefactor is given by $c_s =N \sqrt{(J {+} 1) (2 J {+} 1)/3} \, \gamma_1^{-s}/R $.

The star product with $\FF$ in \eref{exmaplestarprod} and \eref{exampleresult}
can be \emph{approximated} using  \eref{result3approspinweight}.
The approximate actions of $\lp$ and $\lpc$ are then given by
\begin{eqnarray}
\fl  \qquad
\lp \, f
&=
[ \FF + \case{1{-}s}{4J} (\ethadj \FF ) \eth 
- \case{1{+}s}{4J} (\eth \FF ) \ethadj]
\, f 
+ \mathcal{O}(J^{-1}) \\
\fl \qquad \lpc \, f
& =
[ (\FF)^* + \case{1{-}s}{4J} (\eth (\FF)^* ) \ethadj
- \case{1{+}s}{4J} (\ethadj (\FF)^*) \eth]
\, f
+ \mathcal{O}(J^{-1}) .
\end{eqnarray}
Using the star-product approximation from
\eref{generalstarprodapprox},
the actions of $\lp$ and $\lpc$ can be expanded into
\begin{eqnarray}
\fl \;
\lp \, f
& =
[ \FF + \case{1{-}s}{4J} ({\partial}_{\alpha} \FF ) {\partial}_{\alpha^*} 
- \case{1{+}s}{4J} ({\partial}_{\alpha^*} \FF ) {\partial}_{\alpha}]
\, f 
+ \mathcal{O}(J^{-1}) \\
\fl \; \lpc \, f
& =
[ (\FF)^* + \case{1{-}s}{4J} ({\partial}_{\alpha^*} (\FF)^* ) {\partial}_{\alpha}
- \case{1{+}s}{4J} ({\partial}_{\alpha} (\FF)^*) {\partial}_{\alpha^*}]
\, f
+ \mathcal{O}(J^{-1}) .
\end{eqnarray}
Knowing that 	$F_{K(0)} =
c_s \Y_{1,-1}(\Omega)$
with $\eth \Y_{1,-1} = \sqrt{2} \Y_{1,-1}^1$,
$\ethadj \Y_{1,-1} = -\sqrt{2} \Y_{1,-1}^{-1}$, and $(\Y_{1,-1})^*=\Y_{1,1}$,
the action of $\lp$ and $\lpc$ at the point $\Omega_0=0$ is given by 
\begin{eqnarray}
 \fl \qquad \lpn(0) \, f= \lpn \, f & 
= c_s [\Y_{1,-1}  -\sqrt{2} \case{1{-}s}{4J} \Y_{1,-1}^{-1} \eth  - \sqrt{2} \case{1{+}s}{4J} \Y_{1,-1}^{1} \ethadj ]  \, f 
+ \mathcal{O}(J^{-1}) \; \textup{ and}  \\
 \fl \qquad  \lpcn(0) \, f= \lpcn \, f &
= c_s [\Y_{1,1}  +\sqrt{2} \case{1{-}s}{4J} \Y_{1,1}^{1} \ethadj  + \sqrt{2} \case{1{+}s}{4J} \Y_{1,1}^{-1} \eth ]  \, f 
+ \mathcal{O}(J^{-1}),
\end{eqnarray}
which are then used in \eref{Dickestate}-\eref{projectors}.

\section*{References}

\providecommand{\newblock}{}

\end{document}